\documentclass[authoryear]{elsarticle}
\usepackage[framed,autolinebreaks,useliterate]{mcode}
\usepackage{natbib}
\usepackage{amsmath}
\usepackage{rotating}

\usepackage{hyperref}
\usepackage{amssymb}
\usepackage{amsfonts}

\usepackage{pdflscape}
\journal{Neuroimage}
\usepackage{color,transparent}
\usepackage{subfig}
\usepackage{caption}
\usepackage{setspace}
\usepackage{multirow}
\usepackage{array}
\newcommand\MyBox[2]{
  \fbox{\lower0.75cm
    \vbox to 1.7cm{\vfil
      \hbox to 1.7cm{\hfil\parbox{1.4cm}{#1\\#2}\hfil}
      \vfil}%
  }%
}

\begin{document}

\begin{frontmatter}
\title{Statistical power and prediction accuracy in multisite resting-state fMRI connectivity}

\author[a,b]{Christian~Dansereau}
\author[a,c]{Yassine~Benhajali}
\author[d]{Celine~Risterucci}
\author[d]{Emilio~Merlo Pich\corref{cor2}}
\author[a]{Pierre~Orban}
\author[e]{Douglas~Arnold}
\author[a,b]{Pierre~Bellec\corref{cor1}}
\ead{pierre.bellec@criugm.qc.ca}
\cortext[cor1]{Corresponding author: pierre.bellec@criugm.qc.ca}
\address[a]{Centre de Recherche de l'Institut Universitaire de G\'eriatrie de Montr\'eal, Montr\'eal, CA}
\address[b]{Department of Computer Science and Operations Research, University of Montreal, Montreal, CA}
\address[c]{D\'epartement d'anthropologie, Universit\' de Montr\'eal, Montr\'eal, CA}
\address[d]{Clinical Imaging, pRED, F.Hoffman-La Roche, Basel, CH}
\address[e]{NeuroRx inc., Montr\'eal, CA}

\cortext[cor2]{Present address: CNS TAU, Takeda Development Centre Europe, London, UK}

%

\begin{abstract}
Connectivity studies using resting-state functional magnetic resonance imaging are increasingly pooling data acquired at multiple sites. While this may allow investigators to speed up recruitment or increase sample size, multisite studies also potentially introduce systematic biases in connectivity measures across sites. In this work, we measure the inter-site effect in connectivity and its impact on our ability to detect individual and group differences. Our study was based on real, as opposed to simulated, multisite fMRI datasets collected in $N=345$ young, healthy subjects across 8 scanning sites with 3T scanners and heterogeneous scanning protocols, drawn from the 1000 functional connectome project. We first empirically show that typical functional networks were reliably found at the group level in all sites, and that the amplitude of the inter-site effects was small to moderate, with a Cohen's effect size below 0.5 on average across brain connections. We then implemented a series of Monte-Carlo simulations, based on real data, to evaluate the impact of the multisite effects on detection power in statistical tests comparing two groups (with and without the effect) using a general linear model, as well as on the prediction of group labels with a support-vector machine. As a reference, we also implemented the same simulations with fMRI data collected at a single site using an identical sample size. Simulations revealed that using data from heterogeneous sites only slightly decreased our ability to detect changes compared to a monosite study with the GLM, and had a greater impact on prediction accuracy. However, the deleterious effect of multisite data pooling tended to decrease as the total sample size increased, to a point where differences between monosite and multisite simulations were small with $N=120$ subjects. Taken together, our results support the feasibility of multisite studies in rs-fMRI provided the sample size is large enough.
\end{abstract}

\end{frontmatter}

\section*{Highlights}

\begin{itemize}
\item Small to moderate systematic site effects in fMRI connectivity.
\item Small impact of site effects on the detection of group differences for sample size $>100$.
\item Linear regression of the sites prior to multivariate prediction do not improve prediction accuracy.
\end{itemize}

\section{Introduction}

\paragraph{Main objective}
Multisite studies are becoming increasingly common in resting-state functional
magnetic resonance imaging (rs-fMRI). In particular, some consortia have
retrospectively pooled rs-fMRI data from multiple independent studies comparing
clinical cohorts with control groups, e.g. normal controls in the 1000
functional connectome project (FCP) \citep{Biswal2010}, children and adolescents
suffering from attention deficit hyperactivity disorder from the ADHD200
\citep{ADHD200,Fair2012}, individuals diagnosed with autism spectrum disorder in
ABIDE \citep{Nielsen2013}, individuals suffering from schizophrenia
\citep{Cheng2015}, or elderly subjects suffering from mild cognitive impairment
\citep{Tam2015}. The rationale behind such initiatives is to dramatically
increase the sample size at the cost of decreased sample homogeneity. The
systematic variations of connectivity measures derived using different scanners,
called site effects, may decrease the statistical power of group comparisons,
and somewhat mitigate the benefits of having a large sample size
\citep{Brown2011,Jovicich2016}. In this work, our main objective was to
quantitatively assess the impact of site effects on group comparisons in rs-fMRI
connectivity.

\paragraph{Group comparison in rs-fMRI connectivity}
In this work, we focused on the most common measure of individual functional connectivity, which is the Pearson's correlation coefficient between the average rs-fMRI time series of two brain regions. To compare two groups, a general linear model (GLM) is typically used to establish the statistical significance of the difference in average connectivity between the groups. Finally a $p$-value is generated for each connection to quantify the probability that the difference in average connectivity is significantly different from zero \citep{Worsley1995,Yan2013a}. If the estimated $p$-value is smaller than a prescribed tolerable level of false-positive findings (see for more detail Table \ref{confusion_matrix}), generally adjusted for the number of tests performed across connections, say $\alpha=0.001$, then the difference in connectivity is deemed significant. 

\begin{table}[tbp]
\centering
\resizebox{0.35\columnwidth}{!}{%
\begin{tabular}{c >{\bfseries}r @{\hspace{0.7em}}c @{\hspace{0.4em}}c @{\hspace{0.7em}}l}
  \multirow{10}{*}{\parbox{1.1cm}{\bfseries\raggedleft Actual\\ value}} & 
    & \multicolumn{2}{c}{\bfseries Detected value} & \\
  & & \bfseries patho & \bfseries no patho & \\
  & \rotatebox[origin=c]{90}{patho} & \MyBox{True}{Positive} & \MyBox{False}{Negative} &\\[2.4em]
  & \rotatebox[origin=c]{90}{no patho} & \MyBox{\bf False}{\bf Positive} & \MyBox{True}{Negative} &\\

\end{tabular}
}
\caption{Confusion matrix.}
\label{confusion_matrix}
\end{table}

\paragraph{Statistical power in group comparisons at multiple sites}
The statistical power of a group comparison study is the probability of finding
a significant difference, when there is indeed a true difference. A careful study design involves the selection of a sample size that is large enough to reach a set level of statistical power, e.g. $80\%$. In the GLM, the statistical power actually depends on a series of parameters \citep{Desmond2002,Durnez2014}: (1) the sample size (the larger the better); (2) the absolute size of the group difference (the larger the better), and, (3) the intrinsic variability of measurements (the smaller the better) (4) the rejection threshold $\alpha$ for the null hypothesis.

\paragraph{Sources of variability: factors inherent to the scanning protocol}
In a multisite (or multi-protocol) setting, differences in imaging or study parameters may add variance to rs-fMRI measures, e.g. the scanner make and model \citep{Friedman2006,Friedman2008}, repetition time, flip angle, voxel resolution or acquisition volume \citep{Friedman2006a}, experimental design such as eyes-open/eyes-closed \citep{Yan2009}, experiment duration \citep{VanDijk2010}, and scanning environment such as sound attenuation measures \citep{Elliott1999}, or head-motion restraint techniques \citep{Edward2000,VanDijk2012}, amongst others. These parameters can be harmonized to some extent, but differences are unavoidable in large multisite studies. The recent work of \cite{Yan2013a} has indeed demonstrated the presence of significant site effects in rs-fMRI measures in the 1000 FCP. Site effects will increase the variability of measures, and thus decrease statistical power. To the best of our knowledge, it is not yet known how important this decrease in statistical power may be. 

\paragraph{Sources of variability: within-subject}
The relative importance of site effects in rs-fMRI connectivity depends on the amplitude of the many other sources of variance. First, rs-fMRI connectivity only has moderate-to-good test-retest reliability using standard 10-minute imaging protocols \citep{Shehzad2009}, even when using a single scanner and imaging session. Differences in functional connectivity across subjects are also known to correlate with a myriad of behavioural and demographic subject characteristics \citep{Anand2007,Sheline2010,Kilpatrick2006}. Taken together, these sources of variance reflect a fundamental volatility of human physiological signals. 

\paragraph{Sources of variability: factors inherent to the site}
In addition to physiology, some imaging artefacts will vary systematically from session to session, even at a single site. For example, intensity non-uniformities across the brain depend on the positioning of subjects \citep{Caramanos2010}. Room temperature has also been shown to impact MRI measures \citep{Vanhoutte2006}. Given the good consistency of key findings in resting-state connectivity across sites, such as the organization of distributed brain networks \citep{Biswal2010}, it is reasonable to hypothesize that site effects will be small compared to the combination of physiological and within-site imaging variance.

\paragraph{Multivariate analysis}
Another important consideration regarding the impact of site effects on group comparison in rs-fMRI connectivity is the type of method used to identify differences. The concept of statistical power is very well established in the GLM framework, which tests one brain connection at a time (mass univariate testing). However, multivariate methods that combine several or all connectivity values in a single prediction are also widely used and likely affected by the site effects. A popular multivariate technique in rs-fMRI is support-vector machine (SVM) \citep{Cortes1995}. In this approach, the group sample is split into a training set and a test set. The SVM is trained to predict group labels on the training set, and the accuracy of the prediction is evaluated independently on the test set.  The accuracy level of the SVM captures the quality of the prediction of clinical labels from resting-state connectivity, but does not explicitly tell which brain connection is critical for the prediction. The accuracy score can thus be seen as a “separability index” between the individuals of two groups in high dimensional space. Altogether, the objectives and measures of statistical risk for SVM and GLM are quite different. Because SVM has the ability to combine measures across connections, unlike univariate GLM tests, we hypothesized that the GLM and SVM will be impacted differently by site effects. Even though the accuracy is expected to be lower for the multisite than the monosite configuration, it as been shown that the generalizability of a predictive model to unseen sites is greater for models trained on multisite than monosite datasets as shown by \cite{Abraham2016}.

\paragraph{Specific objectives}
Our first objective was to characterize, using real data, the amplitude of
systematic site effects in rs-fMRI connectivity measures across sites, as a function
of within-site variance. We based our evaluation on images generated from
independent groups at 8 sites equipped with 3T scanners, in a subset
($N=345$) of the 1000 FCP. Our second objective was to evaluate the impact of
site effects on the detection power of group differences in rs-fMRI
connectivity. To answer this question directly, one would need to scan two different cohorts of participants at least twice, once in a multisite setting and once in a monosite setting. Such an experiment may be too costly to implement for addressing a purely technical objective. As a more feasible alternative, we implemented a series of Monte Carlo simulations, adding synthetic ``pathological'' effects in the 1000 FCP sample.  One interesting feature of the "1000 FCP" dataset is the presence of one large site of $\sim200$ subjects and 7 small sites of $\sim20$ subjects per site. We were therefore able to implement realistic scenarios following either a monosite or a multisite design (with 7 sites), with the same total sample size. Our simulations gave us full control on critical aspects for the detection of group differences, such as the amplitude of the group difference, sample
size, and the balancing of groups across sites. We evaluated the ability of detecting group differences both in terms of sensitivity for a GLM and in terms of accuracy for a SVM model. 

\section{Method}

\subsection{Imaging sample characteristics}
The full 1000 FCP sample includes 1082 subjects, with images acquired over 33 sites spread across North America, Europe, Australia and China. As the 1000 FCP is a retrospective study, no effort was made to harmonize population characteristics or imaging acquisition parameters \citep{Biswal2010}. A subset of sites was selected based on the following criteria: (1) 3T scanner field strength, (2) full brain coverage for the rs-fMRI scan, and, (3) a minimum of 15 young or middle aged adult participants, with a mixture of males and females (4) samples drawn from a population with a predominant Caucasian ethnicity. In addition, only young and middle aged participants (18-46 years old) were included in the study, and we further excluded subjects with excessive motion (see next Section). The final sample for our study thus included 345 cognitively normal young adults (150 males, age range: 18-46 years, mean$\pm$std: 23.8 $\pm$5.14) with images acquired across 8 sites located in Germany, the United Kingdom, Australia and the United States of America. The total time of available rs-fMRI data for these subjects ranged between 6 and 7.5 min and only one run was available per subject. See Table \ref{table_dataset} for more details on the demographics and imaging parameters at each site selected in the study.  The experimental protocols for all datasets as well as data sharing in the 1000 FCP were approved by the respective ethics committees of each site. This secondary analysis of the 1000 FCP sample was approved by the local ethics committee at CRIUGM, University of Montreal, QC, Canada.

\begin{table}[tbp]
\resizebox{\columnwidth}{!}{%
\begin{tabular}{lclcccccccc}
\textbf{Site}         & \textbf{Magnet } & \textbf{Scanner } & \textbf{Channels } & \textbf{N } & \textbf{Nfinal } & \textbf{Sex } & \textbf{Age } & \textbf{TR } & \textbf{\#Slices } & \textbf{\#Frames} \\
\hline
Baltimore, USA        & 3T              & Philips Achieva       & 8                 & 23   & 21         & 8M/15F       & 20-40        & 2.5         & 47                 & 123                \\
Berlin, DE            & 3T              & Siemens Tim Trio      & 12                & 26   & 26         & 13M/13F      & 23-44        & 2.3         & 34                 & 195                \\
Cambridge, USA        & 3T              & Siemens Tim Trio      & 12                & 198  & 195        & 75M/123F     & 18-30        & 3           & 47                 & 119                \\
Newark, USA           & 3T              & Siemens Allegra       & 12                & 19   & 17         & 9M/10F       & 21-39        & 2           & 32                 & 135                \\
NewYork\_b, USA       & 3T              & Siemens Allegra       & 1                 & 20   & 18         & 8M/12F       & 18-46        & 2           & 33                 & 175                \\
Oxford, UK            & 3T              & Siemens Tim Trio      & 12                & 22   & 20         & 12M/10F      & 20-35        & 2           & 34                 & 175                \\
Queensland, AU        & 3T              & Bruker                & 1                 & 19   & 17         & 11M/8F       & 20-34        & 2.1         & 36                 & 190                \\
SaintLouis, USA       & 3T              & Siemens Tim Trio      & 12                & 31   & 31         & 14M/17F      & 21-29        & 2.5         & 32                 & 127              
\end{tabular}%
}
\caption{Sites selected from the 1000 Functional Connectome Project.}
\label{table_dataset}
\end{table}

\subsection{Computational environment}
All experiments were performed using the NeuroImaging Analysis Kit, NIAK\footnote{\url{http://simexp.github.io/niak/}} \citep{Bellec2011} version 0.12.18, under CentOS version 6.3 with Octave\footnote{\url{http://gnu.octave.org/}} version 3.8.1 and the Minc toolkit\footnote{\url{http://www.bic.mni.mcgill.ca/ServicesSoftware/ServicesSoftwareMincToolKit}} version 0.3.18. Analyses were executed in parallel on the ”Mammouth” supercomputer\footnote{\url{http://www.calculquebec.ca/index.php/en/resources/compute-servers/mammouth-serie-ii}} , using the pipeline system for Octave and Matlab, PSOM \citep{Bellec2012} version 1.0.2. The scripts used for processing can be found on Github\footnote{\url{https://github.com/SIMEXP/Projects/tree/master/multisite}}. Prediction was performed using the LibSVM library \citep{Chang2011}. Visualization was implemented using Python 2.7.9 from the Anaconda 2.2.0\footnote{\url{http://docs.continuum.io/anaconda/index}} distribution, along with Matplotlib\footnote{\url{http://matplotlib.org/}} \citep{matplotlib}, Seaborn\footnote{\url{http://stanford.edu/~mwaskom/software/seaborn/index.html}} and Nilearn\footnote{\url{http://nilearn.github.io/}} for brain map visualizations.

\subsection{Preprocessing}
Each fMRI dataset was corrected for slice timing; a rigid-body motion was then estimated for each time frame, both within and between runs, as well as between one fMRI run and the T1 scan for each subject \citep{Collins1994}. The T1 scan was itself non-linearly co-registered to the Montreal Neurological Institute (MNI) ICBM152 stereotaxic symmetric template \citep{Fonov2011}, using the CIVET pipeline \citep{Ad-Dab'bagh2006}. The rigid-body, fMRI-to-T1 and T1-to-stereotaxic transformations were all combined to re-sample the fMRI in MNI space at a 3 mm isotropic resolution. To minimize artifacts due to excessive motion, all time frames showing a frame displacement, as defined in \cite{Power2012}, greater than 0.5 mm were removed and a residual motion estimated after scrubbing. A minimum of 50 unscrubbed volumes per run was required for further analysis (13 subjects were rejected). The following nuisance covariates were regressed out from fMRI time series: slow time drifts (basis of discrete cosines with a 0.01 Hz highpass cut-off), average signals in conservative masks of the white matter and the lateral ventricles (average Pearson correlation across all subjects is 0.242  between gray matter and white matter signals, and 0.031 between gray matter and ventricles signals) as well as the first principal components (accounting for 95\% variance) of the six rigid-body motion parameters and their squares \citep{Giove2009,Lund2006}. The fMRI volumes were finally spatially smoothed with a 6 mm isotropic Gaussian blurring kernel. A more detailed description of the pipeline can be found on the NIAK website\footnote{\url{http://niak.simexp-lab.org/pipe_preprocessing.html}} and Github\footnote{\url{https://github.com/SIMEXP/}}.

\subsection{Inter-site bias in resting-state connectivity}
\paragraph{Functional connectomes} We compared the functional connectivity measures derived from different sites of the 1000 FCP. A functional brain parcellation with $100$ regions was first generated using a bootstrap analysis of stable clusters \citep{Bellec2010c}, on the Cambridge cohort of the 1000 FCP ($N=195$), as described in \cite{Orban2015}. For a given pair of regions, the connectivity measure was defined by the Fisher transformation of the Pearson's correlation coefficient between the average temporal rs-fMRI fluctuations of the two regions. For each subject, a $100 \times 100$ functional connectome matrix was thus generated, featuring the connections for every possible pair of brain regions. 

\paragraph{Inter-site effects} The inter-site effects at a particular connection were defined as the absolute difference in average connectivity between two sites. In order to formally test the significance of the inter-site effects, we used a GLM including age, sex and residual motion as covariates (corrected to have a zero mean across subjects), as well as dummy variables coding for the average connectivity at each site. For each site, a ``contrast'' vector was coded to measure the difference in average connectivity between this site and the grand average of functional connectivity combining all other sites. A $p$-value was generated for each connection to quantify the probability that the observed effect using this contrast was significantly different from zero \citep{Worsley1995}. The number of false discovery was also controlled ($q=0.05$) using a Benjamini-Hochberg false discovery rate (FDR) procedure \citep{Benjamini1995}.  To quantify the severity of inter-site effects, we derived Cohen's $d$ effect size measure for each connection: $|\beta_c|/\hat{\sigma}$, with $\beta_c$ being the weight associated with the contrast. The standard
deviation from the noise $\hat{\sigma}$ was calculated as $\hat{\sigma}=\sqrt{\sum{e^{2}}/(N-K)}$, $e$ being the residuals from the GLM, $N$
the sample size and $K$ the number of covariates in the model. As secondary analyses, $t$-tests were also implemented in the GLM to validate that age, sex as well as residual motion made significant contributions to the model. 

\subsection{Simulations}

\paragraph{Data generation process}
We implemented Monte-Carlo simulations to assess the detection sensitivity of group differences in rs-fMRI connectivity. The simulations were based on the 1000 FCP sample, with 8 sites totaling $345$ subjects. The multisite simulations were sampled from $148$ subjects, available across $S=7$ sites. The monosite simulations were sampled from $195$ subjects available at $S=1$ site (Cambridge). For each simulation, a subset of subjects of a given size $N$ was selected randomly and stratified by site. For each site, a ratio $W$ of the selected subjects was randomly assigned to a so-called  ``patient'' group. 
We focus our analysis on connections showing a fair-to-good test-retest reliability based on a previous study reporting 11 connections likely impacted by Alzheimer's disease, see \cite{Orban2015} for details. For each connection, a ``pathology'' effect was added to the connectivity measures of the subjects belonging to the ``patient'' group. This additive shift in connectivity for ``patients'' was selected as to achieve a specified effect size, defined below.

\paragraph{Effect size (Cohen's $d$)}
The Cohen's $d$ was used to quantify the effect size. For a group comparison, Cohen's $d$ is defined as the difference $\mu$ between the means of the two groups, divided by the standard deviation of the measures within each group, here assumed to be equal. For a given connection between brain regions $i$ and $j$, let $y_{i,j}$ be the functional connectivity measure for a particular subject of the 1000 FCP sample. If the subject was assigned to the ``patient'' group in a particular simulation, an effect was added to generate a simulated connectivity measure $y_{i,j}^*$ equal to $y_{i,j} + \mu$. For a specified effect size $d$, the parameter $\mu$ was set to $d\times s_{i,j}$, where $s_{i,j}$ is the standard deviation of connectivity between region $i$ and $j$. The parameter $s_{i,j}$ was estimated as the standard deviation of connectivity measures across subjects in the mono-site sample (Cambridge), without any ``pathological'' effect simulated.

\paragraph{GLM tests}
In order to detect changes between the simulated groups at each connection, a GLM was estimated from the simulated data, using age, sex and frame displacement as confounds (corrected to have a zero mean across subjects). To account for site-specific effects, $S-1$ dummy variables (binary vectors coding for each site) were added to the model, with $S$ being the total number of sites used in the study, in addition to an intercept accounting for the global average. Finally, one dummy variable coded for the ``patient'' group. The regression coefficients of the linear model were estimated with ordinary least squares, and a $t$-test, with associated $p$-value, was calculated for the coefficient of the ``patient'' variable. A significant pathology effect was detected if the $p$ value was smaller than a prescribed $\alpha$ level. The $\alpha$ level needs to be adjusted for multiple comparisons (in our case 11 connections, but this would depend on the number of connections selected in a particular study), which can be done in an adaptive manner using FDR. When connections are pre-specified, such as in e.g. \cite{Wang2012}, a more liberal threshold can be applied. In our case, since we wanted to have a constant behavior independent of the effect size, we tested different typical values for $\alpha$ in $\{0.001,0.01,0.05\}$. For each simulation sample $b$ and each connection, we derived a $p$-value $p^{(*b)}$, and the effect was deemed detected if $p^{*b}$ was less than $\alpha$. The sensitivity of the test for a particular connection was evaluated by the frequency of positive detections over all simulation samples.

\paragraph{Prediction accuracy}
In addition to mass univariate GLM tests, we also investigated a linear SVM
\citep{Cortes1995} using a Monte-Carlo simulation of the prediction of clinical labels based on cross-validation. For SVM simulations, all possible connections between the 100
brain regions were used simultaneously to predict the presence of the simulated
pathology in a given subject. For a participant assigned to the ``patient''
group, a ``pathology'' effect was only simulated in a set percentage of
connections, which were randomly selected. The proportion of connections with a
non-null effect was denoted as $\pi_1$. For a given simulation at sample size
$N$, the SVM model was trained on $N$ subjects selected randomly and stratified
by site. The accuracy of the model was evaluated on a separate sample consisting of the remaining subjects, unused during training. For example, for a multisite simulation with $N=80$ subjects for training, the model accuracy was tested on the remaining $68$ subjects: $148$ (available subjects) minus $80$ (subjects in the training set). During training, a 10-fold cross-validation was used to optimize the hyper-parameters of the SVM independently for each
simulation. The mean and standard deviation of accuracy scores across all
samples were derived for each simulation scenario.

\paragraph{Simulation experiments}

All the simulation parameters have been summarized below: 
\begin{itemize}
 \item Sample size $N$. 
 \item Patient allocation ratio $W$. 
 \item Number of sites $S$. 
 \item The type of detection method, either GLM or SVM. 
 \item For GLM tests, the false-positive rate $\alpha$. 
 \item For SVM tests, the proportion of ``pathological'' connections $\pi_1$.
 \item The effect size $d$. 
\end{itemize}

For a given set of simulation parameters, we generated $B=10^3$ Monte-Carlo
samples to estimate either the sensitivity (for GLM test) or the accuracy (for
SVM prediction) of the method. For all experiments, we investigated effect sizes
$d\in\{0,2\}$ with a step of $0.01$ and $\alpha\in\{0.001,0.01,0.05\}$. The
number of site(s) was $S=1$ for the monosite analysis and $S=7$ for the
multisite analysis. We implemented the following experiments:

\begin{itemize}
\item $(\mathcal{E}_1)$ Test the impact of the sample size on GLM
$N\in\{40,80,120\}$, with a fixed allocation ratio $W=0.5$.
\item $(\mathcal{E}_2)$ Test the impact of the allocation ratio on GLM
$W\in\{0.5,0.3,0.15\}$ for a fixed sample size $N=120$.
\item $(\mathcal{E}_3)$ Test the impact of multisite correction (regressing out the site effects using dummy variables coding for each site) and affected
connection volume ($\pi_1$) on the prediction accuracy. For the prediction
scenario, we used a range of $\pi_1\in\{0.1,1,5\%\}$, and two sample sizes
$N\in\{80,120\}$ subjects for training, with model accuracy estimated on $N=68$
and $N=28$, respectively.
\end{itemize}

\section{Results}

\subsection{Inter-site effects in fMRI connectivity}

\begin{figure}[htbp]
\begin{center}
\includegraphics[width=0.8\linewidth]{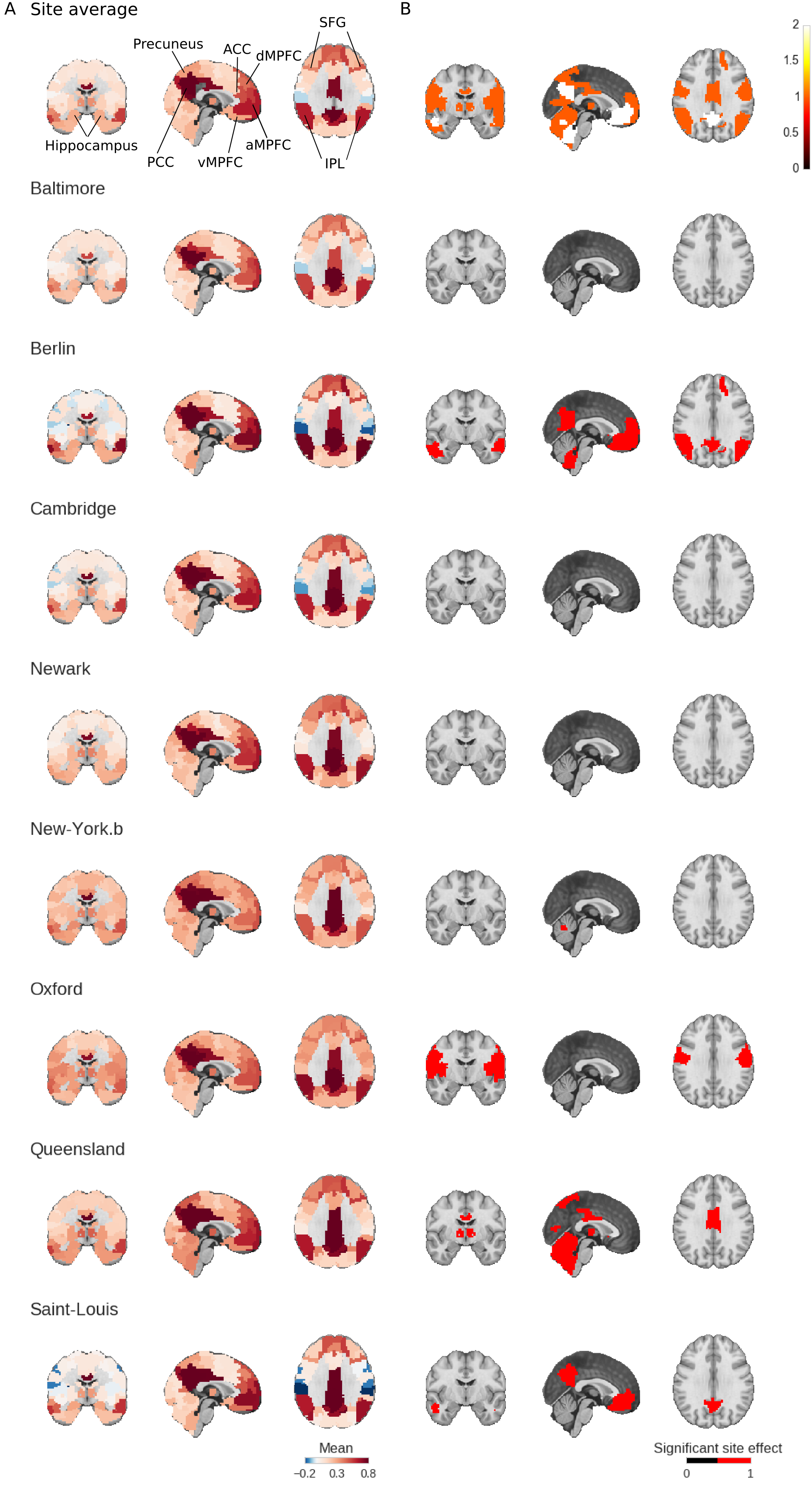}
\end{center}
\caption[DMN variability across sites]{
Panel A: map of the DMN obtained using a seed in the posterior cingulate cortex, averaging all subjects and sites together (first row) and then averaging all subjects for each of the 8 sites (subsequent rows). Panel B shows the number of sites with a significant inter-site difference for each brain region (first row) and the significant differences between the average functional connectivity maps of one site versus all the others (subsequent rows).
}
\label{fig_DMN_variability}
\end{figure}

\paragraph{Site effects in the default-mode network} We first focused on the
connections associated with a seed region located in the posterior cingulate cortex, a key
node of the default-mode network (DMN), which is one of the most widely studied
resting-state networks \citep{Greicius2004}. The connections were based on the
Cambridge $100$ parcellation, and were represented as a connectivity map, (Figure
\ref{fig_DMN_variability}). Figure \ref{fig_DMN_variability}A shows the posterior cingulate cortex
connectivity map, averaged across all subjects and all sites. The key regions of
the DMN are easily identifiable, and include the posterior cingulate cortex, precuneus, inferior
parietal lobule, anterior cingulate cortex, medial pre-frontal cortex (dorsal,
anterior and ventral), superior frontal gyri and the medial temporal lobe
\citep{Damoiseaux2006,Dansereau2014,Yan2013a}. The average
connectivity map of the DMN was then extracted for each site, Figure
\ref{fig_DMN_variability}A. Qualitatively, the DMN maps were consistent across
sites, as expected based on the literature. We then tested for the significance
of the site effects (Figure \ref{fig_DMN_variability}B), i.e. the difference in average connectivity at a given site
and the average connectivity at all remaining sites. The statistical maps were
corrected for multiple comparisons across the brain with FDR at $q\leq 0.05$
\citep{Benjamini1995}. A significant site effect for at least one connection could be
identified for every site, without exception, Figure
\ref{fig_DMN_variability}B. Figure \ref{fig_DMN_variability}C shows how
reproducible the significant site effects were in connectivity across the brain and
sites. The identified significant connections were quite variable across sites, most of them being
identified at less than three sites.

\begin{figure}[htbp]
\begin{center}
\includegraphics[width=\linewidth]{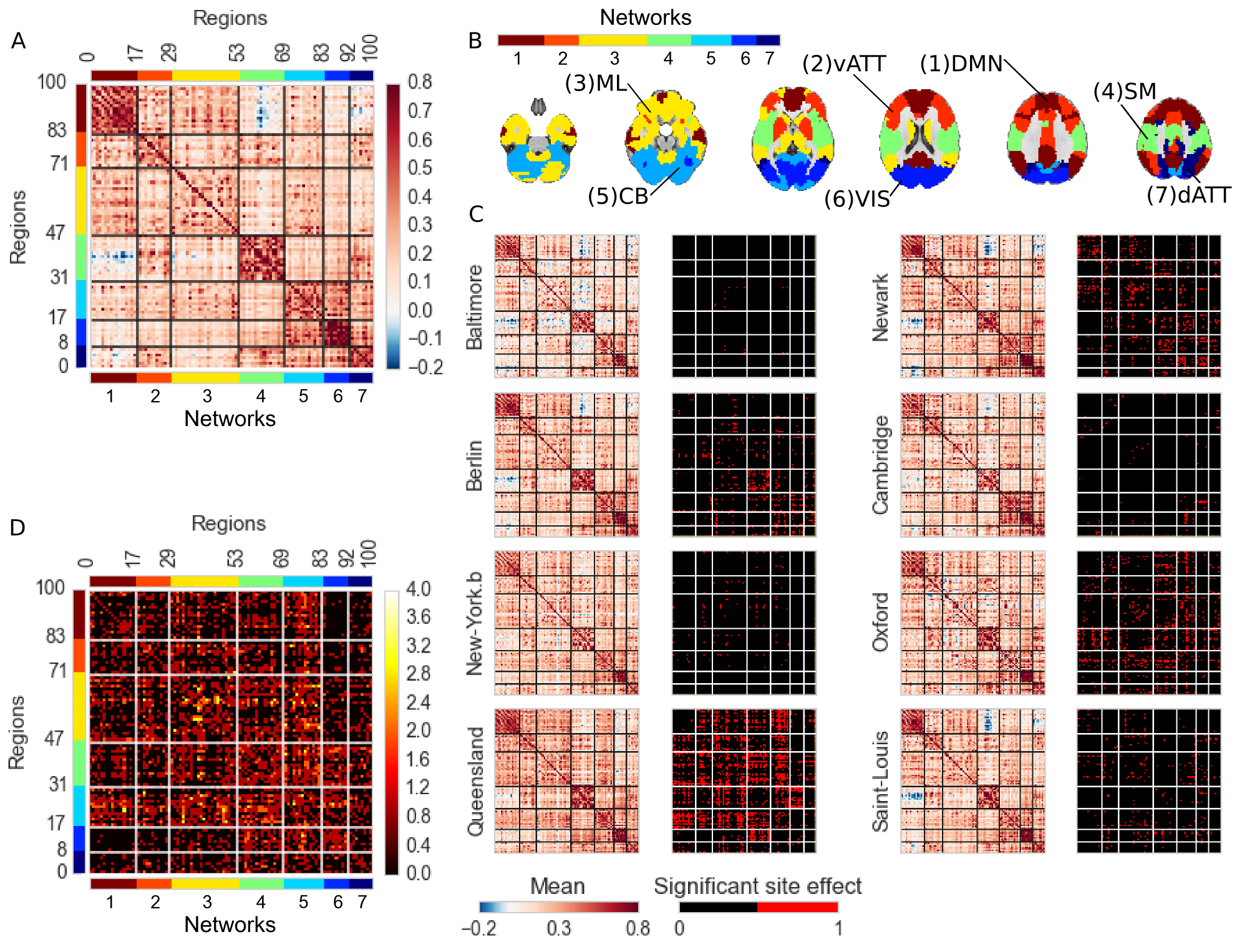}
\end{center}
\caption[Connectome variability across sites]{
Panel A shows the average functional connectomes for 8 sites of the 1000 FCP. Colors next to the $x$ and $y$ axis correspond to different networks in a 7-cluster solution of the matrix, obtained from a hierarchical clustering (Ward criterion). Panel B presents the corresponding 7 brain networks, along with labels. Panel C shows average connectomes for individual sites, as well as connections with a significant site effect. Panel D shows the number of sites at which a given connection was detected as significant. ML: mesolimbic, CB: cerebellar, VIS: visual, vATT: ventral attentional, dATT: dorsal attentional, DMN: default mode network, SM: sensorimotor.
}
\label{fig_connectome_variability}
\end{figure}

\paragraph{Site effects across the connectome} In order to extend these
observations outside of the DMN, we derived the entire connectome using the
Cambridge 100 parcellation. Figure \ref{fig_connectome_variability}A shows the
average connectome, pooling all subjects and sites together. The regions have
been re-ordered based on a hierarchical clustering with Ward criterion. A
network structure is clearly visible as squares of high connectivity on the
diagonal of the connectome (as outlined by black lines). Each diagonal square
corresponds to the intra-network connectivity for a partition into 7 networks (Figure \ref{fig_connectome_variability}A). These 7 networks\footnote{\url{http://neurovault.org/images/39184/}} were
consistent with the major resting-state networks reported using a cluster
analysis in previous works \citep[e.g.][]{Heuvel2008, Bellec2010, Yeo2011,
Power2011}: the DMN, visual, sensorimotor, dorsal and ventral attentional
networks, mesolimbic and cerebellar networks were
identified (Figure \ref{fig_connectome_variability}B). Figure \ref{fig_connectome_variability}C shows how this large-scale
connectome organization varied from site to site. The average connectivity per
site as well as significant differences with the average of the remaining sites
($q\leq 0.05$) is shown in Figure \ref{fig_connectome_variability}C.
Visually, consistent with our previous observations in the DMN, the organization
of the average connectome into large-scale resting-state networks was preserved
across all sites. 

Some significant site effects were still detected in the connectivity both within each network, as well as between networks. By counting the number of sites showing a significant effect for each pair of regions, it was apparent that significant site effects were quite variable in their localization and spread across the full connectome (Figure \ref{fig_connectome_variability}D). 
Concerning the association with the other confounding variables in the model (sex, age and motion) many connections were found to be significantly associated with motion, see Supplementary Material Figure \ref{fig_connectome_variability_motion}, although very few connections were found to be significantly associated with the sex and age, see Supplementary Material Figure \ref{fig_connectome_variability_sex} and \ref{fig_connectome_variability_age}. We also checked that the analysis was not predominantly driven by the larger Cambridge site. We thus ran the same analysis excluding that site (see Supplementary Material Figure \ref{fig_connectome_variability_no_cambridge}). The number of significant pairs remained very similar, although the spatial location of half of the significant connectivity pairs changed when the large Cambridge site was removed from the analysis. Those findings do not qualitatively change our conclusion, but they influence the location of the significant connections. These differences may be due to the intrinsic variability in the statistical test, and not just the size of the Cambridge site. In summary, those findings support the inclusion of age, sex and motion parameters in a GLM in order to remove their confounding effects in addition to site effects.

\begin{figure}[htbp]
\begin{center}
\includegraphics[width=0.8\linewidth]{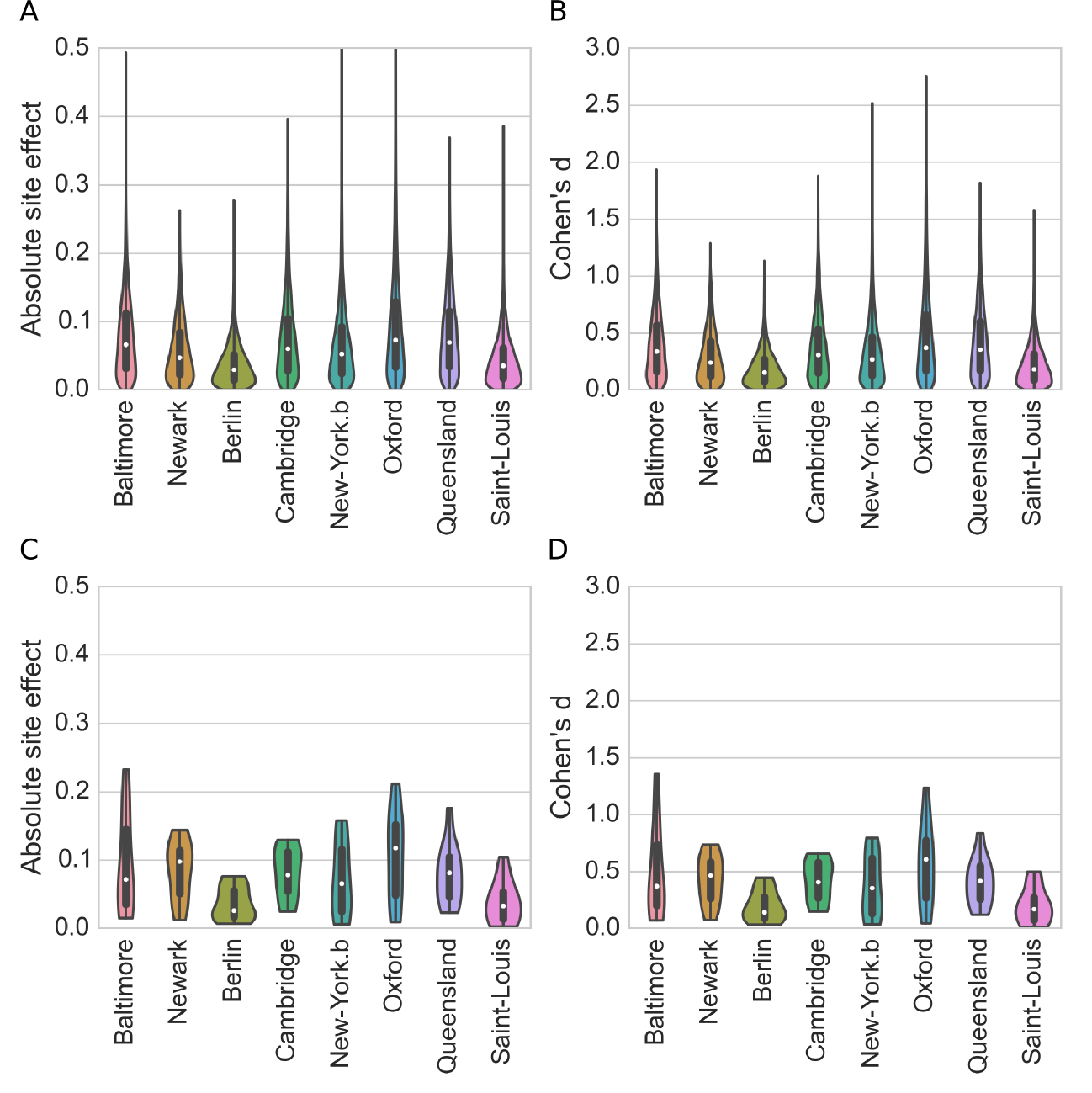}
\end{center}
\caption[inter site variability]{
Effect size of the inter-site effects from a subset of 8 sites from the 1000 FCP. Panels A,C show the distribution of absolute differences in functional connectivity, while panels B,D show Cohen's $d$ measures of inter-site effects. Panels A,B show violin plots across every connections in the BASC Cambridge 100 parcellation, while Panels C,D focus on the selected 11 functional connections used in simulations, only. 
}
\label{fig_site_variability}
\end{figure}

\paragraph{Site effects vs. within-site variations across subjects} We measured the amplitude of inter-site effects, represented as violin plots across connections using either the absolute difference in average connectivity (Figure \ref{fig_site_variability}A,C) or Cohen's $d$ effect size measures (Figure \ref{fig_site_variability}B,D). The violin plots include either every connection from the BASC Cambridge parcellation (Figure \ref{fig_site_variability}A,B), or only the 11 connections selected for Monte-Carlo
simulations (Figure \ref{fig_site_variability}C,D). For absolute differences, the distributions were mostly consistent across sites, with a
median around 0.06, $5\%$ percentile near 0 and $95\%$ percentiles in the 0.08-
0.1 range. For Cohen's $d$, the distributions were also consistent across sites, with a
median around 0.33, $5\%$ percentile near 0 and $95\%$ percentiles in the 0.4-
0.6 range. These effect sizes are typically deemed small-to-moderate \citep{Cohen1992}, although such a qualitative assessment needs to be refined based on each application. This result thus suggests that the impact of additive inter-site effects on statistical tests will be limited. Similar findings were observed across all possible connections, or across the 11 pairs of
connections selected in the simulation study. 

\paragraph{Differences in standard deviation across sites} We also investigated the site differences in standard
deviation of connectivity across subjects, see Supplementary Figure \ref{fig_std_DMNs} for the DMN,
Supplementary Material
\ref{fig_std_connectomes} for the connectomes. The standard group GLM assumes equal variance of resting-state connectivity across all subjects, or ``homoscedasticity''. Significant differences in across-subject standard deviation between sites violates the homoscedastic assumption, and may jeopardize the validity of the false-positive rates of the model. Qualitatively, we first observed that the sites showing the larger number of differences were the one with the most temporal variance among connections see Supplementary Figure \ref{fig_std_ts_distribution}. We then ran a White's test aimed at rejecting homoscedasticity at each connection, independently. The White's tests resulted in a family of p-values, which was corrected for multiple comparisons using FDR ($q<0.05$). The homoscedastic hypothesis was rejected in a large portion of connections. This was expected due to the large overall number of subjects and consequently large statistical power of White's procedure. However, despite reaching significance, the absolute difference in the average standard deviation between two sites was 19\% of the grand average standard deviation, on average across pairs of sites. Such a small departure from homoscedasticity likely has only a mild impact on the GLM, which we formally investigated using Monte-Carlo simulations. 

\subsection{Multisite Monte-Carlo simulations}

\begin{figure}[htbp]
\centering
{\includegraphics[width=\textwidth]{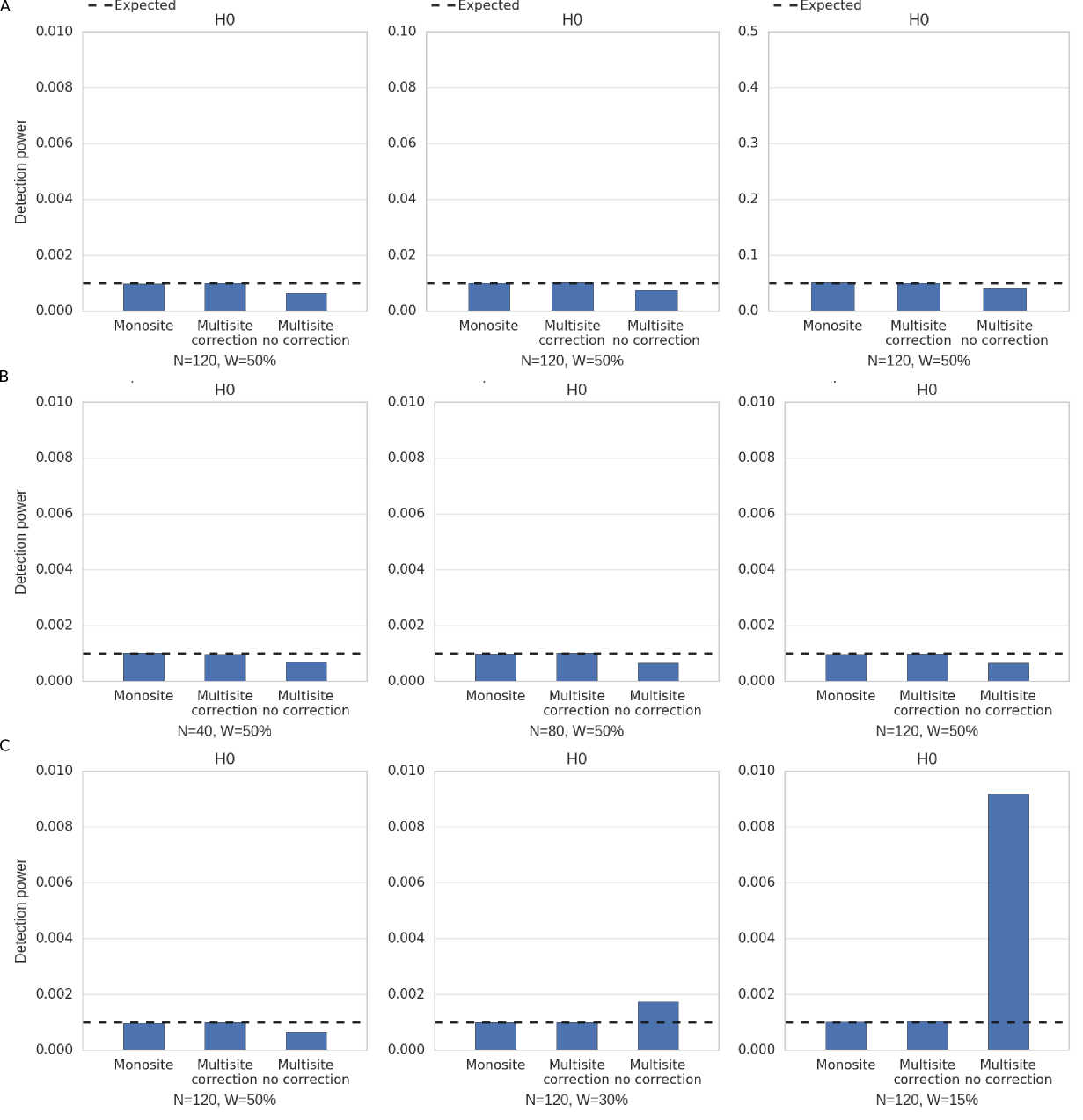}}

\caption{
Monte-Carlo simulation of the false positive rate in the absence of group differences ($d=0$), either for a monosite ($S=1$, left), a multisite ($S=7$) with (middle) or without (right) site covariates included in the GLM. In panel A, three different $\alpha$ values have been tested, $\alpha\in \{0.001, 0.01, 0.05\}$ with a fixed sample size and patient allocation ratio ($N=120,W=50\%$). In panel B, three different sample sizes have been tested, $N\in \{40, 80, 120\}$ with a fixed patient allocation ratio ($W=50\%$) (Experiment $(\mathcal{E}_1)$). In panel C, three different patient allocation ratios have been tested, $W\in \{50\%, 30\%, 15\%\}$ with a fixed sample size ($N=120$) (Experiment $(\mathcal{E}_2)$).}
\label{fig_h0}
\end{figure}

\paragraph{Validity of the control of false positives in the GLM} An excellent control of the false positive rate was observed at all nominal levels $\alpha\in \{0.001, 0.01, 0.05\}$, both in monosite simulations or in multisite simulations, when site covariates were included in the GLM, see Figure~\ref{fig_h0}. This means that the nominal, user-specified, false positive rate matched precisely with the effective false positive rate measured in the simulations. This observation held for any combination of allocation ratio, $W\in\{15\%,30\%,50\%$\}, and sample size, $N\in\{40, 80, 120\}$. By contrast, when no site covariates were included in the GLM, the false positive rate was not controlled appropriately, sometimes by a wide margin. In the absence of site covariates, the procedure was sometimes too conservative, e.g. $W=50\%$, and sometimes very liberal, e.g. $N=120, W=15\%$. This experiment showed that, despite the mild departure from homoscedasticity reported above, the GLM does control for false-positive rate at each connection very precisely, if and only if site covariates are included in the model.

\begin{figure}[htbp]
\centering
{\includegraphics[width=\textwidth]{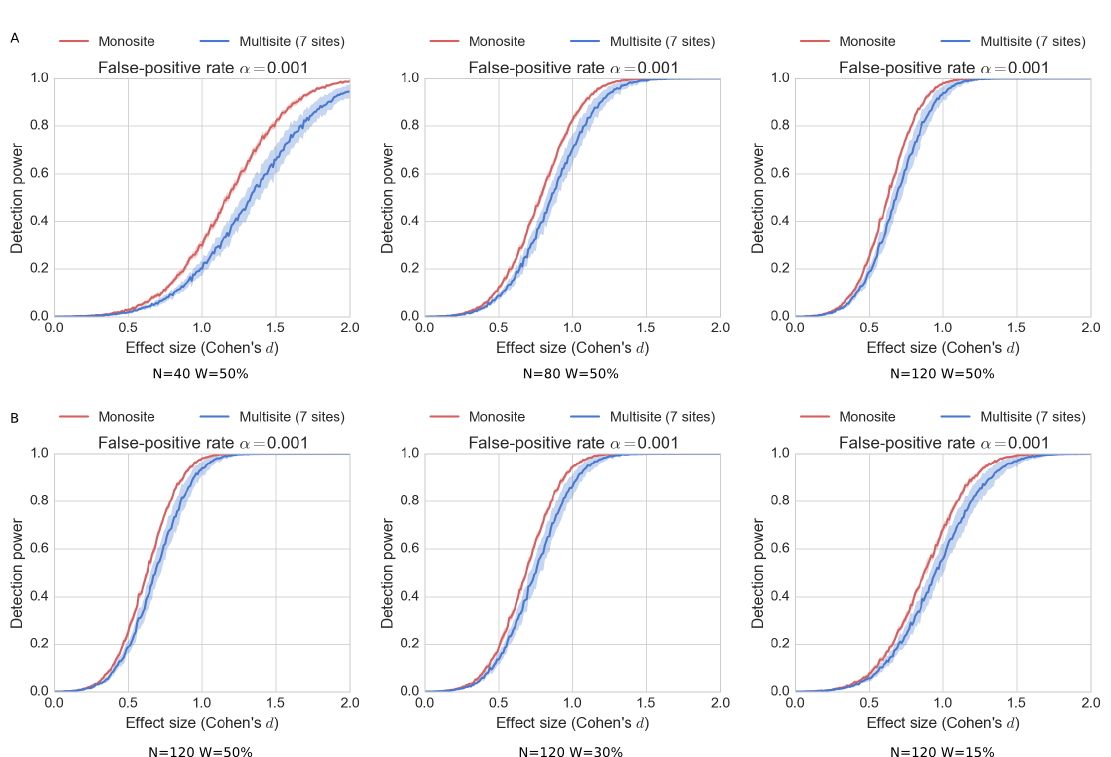}}

\caption{
Monte-Carlo simulation of detection power as a function of the effect size $d\in[0,2]$, either for a monosite ($S=1$, in red) or a multisite ($S=7$, in blue) sample, when testing differences between two groups with a GLM and a false-positive rate $\alpha=0.001$. The plain curves are the average statistical power across 11 connections, and the shaded area represents $\pm 1$ standard deviation across connections. In panel A, the patient allocation ratio is fixed ($W=50\%$) and three different sample sizes have been tested, $N\in\{40, 80, 120\}$ (Experiment $(\mathcal{E}_1)$). In panel B, the sample size is fixed ($N=120$) and three different patient allocation ratios have been tested $W\in \{15\%, 30\%,50\%\}$ (Experiment $(\mathcal{E}_2)$).
}
\label{fig_real_sim}
\end{figure}

\paragraph{Statistical power and effect size} Figure \ref{fig_real_sim}A shows the relationship between effect size and a GLM detection power in experiment $(\mathcal{E}_1)$, i.e. for a fixed allocation ratio ($W=50\%$) and three different sample sizes, $N\in\{40, 80, 120\}$. The average and std of detection power was plotted across the 11 selected connections. The variations of statistical power across connections were very small for monosite simulations, as the effect size was adjusted based on the standard deviation of each connection within that sample. As expected, the sensitivity increased with sample size, quite markedly. In multisite simulations ($S=7$), for a large effect size ($d=1$), the detection power was $20\%$ with 40 subjects , $80\%$ with 80 subjects and $95\%$ with $120$ subjects. The sensitivity was larger with a single site than a multisite sample, yet the difference between the two decreased as sample size increased. With $N=40$ and $d=1$, the detection power was close to $30\%$ for a single site sample, compared to $20\%$ for the multisite sample. With $N=120$ and $d=1$, the difference in sensitivity was only of a few percent. The same trend was apparent for all tested effect sizes as well as for $\alpha\in\{0.01,0.05\}$ (not shown).
\par
\begin{figure}[htbp]\centering
\includegraphics[width=\textwidth]{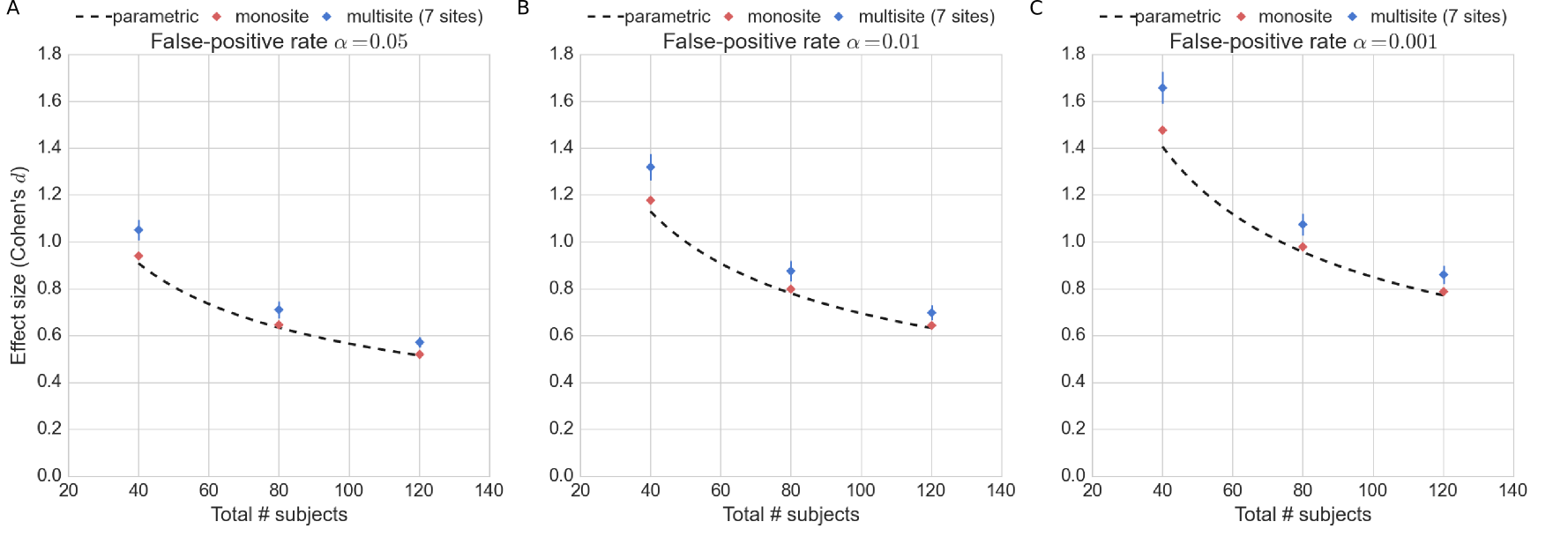}
\caption[]{
Effect size detectable at $80\%$ sensitivity as a function of sample size, for different false-positive rate $\alpha\in \{0.05,0.01,0.001\}$ (experiment $(\mathcal{E}_1)$). All simulations used a balanced patient allocation ratio $W=50\%$. The monosite performance is shown in red and the multisite in blue. The dotted black line shows the detectable effect size for a classical parametric $t$-test. 
}
\label{fig_sampeffect_curves_alpha001}
\end{figure}

\paragraph{Statistical power and group allocation ratio} Figure \ref{fig_real_sim}B shows the relationship between effect size and a GLM detection power in experiment $(\mathcal{E}_2)$, i.e. for a fixed sample size ($N=120$) and three different patient allocation ratio, $W\in \{15\%, 30\%,50\%\}$. Overall, we found that the detection power increased with $W$. For example, with $d=1$, the detection power was $65\%$ for $W=15\%$, and increased to $90\%$ with $W=30\%$, and finally $95\%$ for $W=50\%$. The impact of $W$ was observed in both monosite and multisite samples, with an optimal allocation ratio of $W=50\%$ for both. This observation was also made for $\alpha\in\{0.01,0.05\}$ (not shown). 

\paragraph{Detectable effect size, as a function of sample size} An alternative summary of experiment $(\mathcal{E}_1)$ is to represent the effect size that can be detected with 80\% sensitivity, as a function of sample size for monosite and multisite configurations, see Figure \ref{fig_sampeffect_curves_alpha001}. As a reference, we computed the same curve for parametric $t$-test comparisons, under assumptions of normality. As expected, the detectable effect size for parametric $t$-tests closely followed the monosite estimation. For a small sample size ($N=40$), the detectable effect size was notably larger in multisite configurations than in a monosite configuration (difference of about 0.25 in Cohen's $d$ for $\alpha=0.001$). However, the difference decreased for large sample sizes to become smaller than 0.1 with $N=120$ and $\alpha=0.001$. The lowest detectable effect size for a sensitivity of $80\%$ at $\alpha=0.05$ was about $d=0.8$, achieved in a monosite configuration with $N=120$. At this sample size, the difference between single and multisite configurations was marginal, with only a few percent's of difference in detectable effect sizes.

\begin{figure}[htbp]
\centering
\captionsetup[subfloat]{labelformat=empty}
\includegraphics[width=\textwidth]{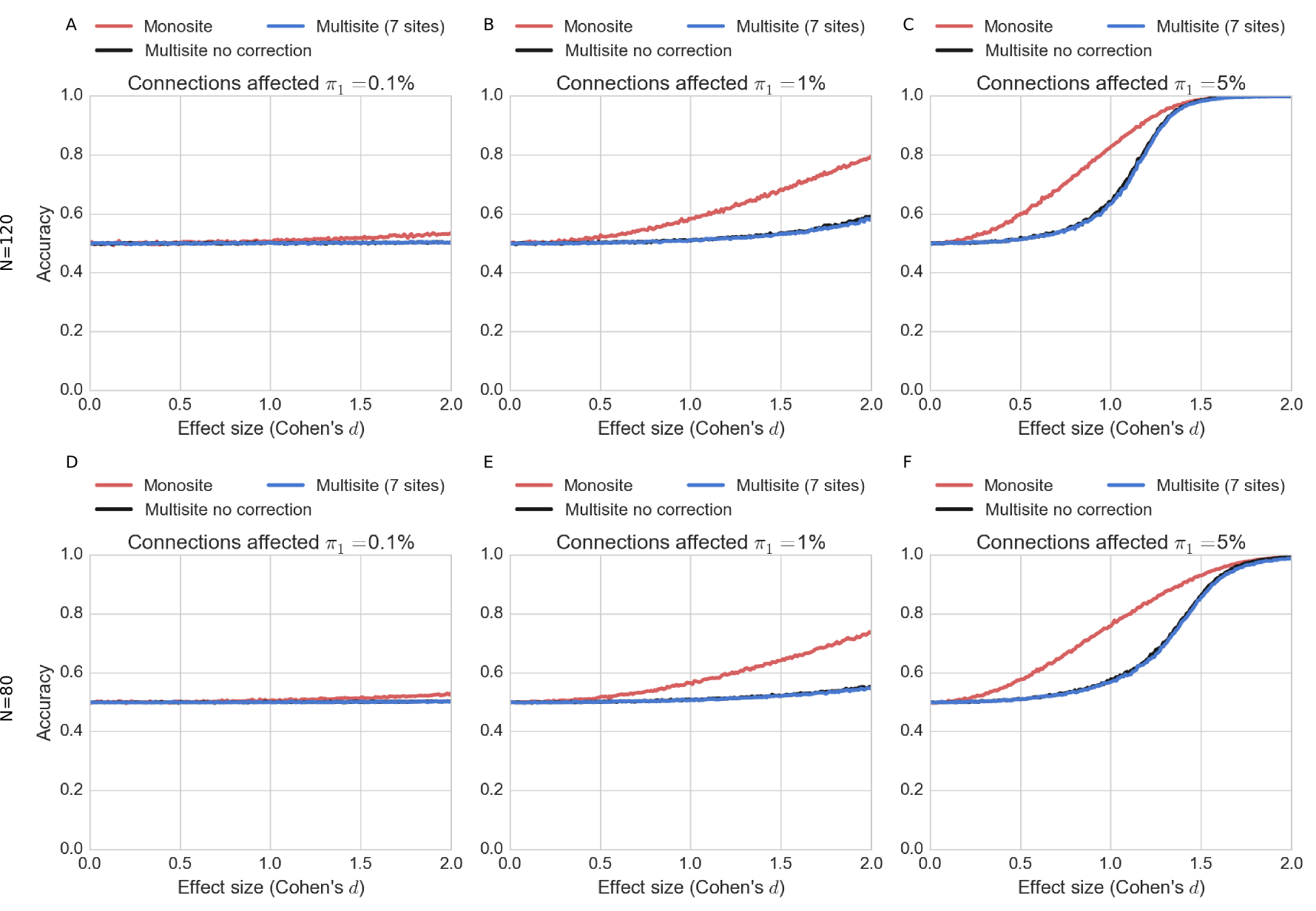}
\caption[]{
Prediction accuracy of patient vs. controls as a function of effect size. Three simulation settings are presented on each plot: monosite (red curve), multisite with regression of site effects ($S=7$, blue curve), and multisite without regression of site effects ($S=7$, black curve). Accuracy was estimated over $B=10^3$ simulation samples with a patient allocation ratio $W=50\%$ and 3 volumes of affected connections $\pi_1=0.1\%$ (left column), $\pi_1=1\%$ (middle column) and $\pi_1=5\%$ (right column). Two sample sizes were tested: $N=120$ randomly selected subjects for training, with the remaining $N=28$ to estimate accuracy (first row), and $N=80$ randomly selected subjects for training, with the remaining $N=68$ to estimate accuracy (second row).
}
\label{fig_prediction_sampeffect}
\end{figure}

\paragraph{Prediction accuracy}
In experiment $(\mathcal{E}_3)$, we examined the impact of effect size and the volume of affected connections on prediction accuracy in a SVM, see Figure \ref{fig_prediction_sampeffect}. The volume of changes $\pi_1$ had a major impact on prediction accuracy. At $\pi_1=0.1\%$ (around 5 connections) the accuracy level was at chance level across all tested effect sizes, (Figure \ref{fig_prediction_sampeffect}A). With $\pi_1=1\%$, accuracy slightly increased, but effect sizes larger than $d=2$ were still required to reach over $80\%$ accuracy (Figure \ref{fig_prediction_sampeffect}B). With $\pi_1=5\%$, $95\%$ accuracy was achieved at the same effect size (about $d=1.5$) for monosite and multisite simulations, although the accuracy in multisite simulations was notably lower than for monosite simulations across most effect sizes (Figure \ref{fig_prediction_sampeffect}C). The relationship between effect size and accuracy followed a sigmoidal curve in both settings, yet a sharper, and later transition between very low and very high accuracy was observed in multisite simulations. Interestingly, correcting for site effects by regressing out the dummy variable before running the SVM classifier had no impact on accuracy levels. The sample size ($N=80$ vs $N=120$ for training) did have a moderate effect on prediction accuracy: for $\pi_1=5\%$ and $d=1$ and monosite simulations, accuracy was about $85\%$ with $N=120$ (Figure \ref{fig_prediction_sampeffect}C) and $75\%$ with $N=80$ (Figure \ref{fig_prediction_sampeffect}F).


\section{Discussion and conclusions}

\paragraph{Inter-site effects in rs-fMRI connectivity} Typical resting-state networks, such as the DMN, the attentional, visual and sensorimotor networks, were reliably found across sites. This was strongly expected given the relative consistency of their distribution across individuals, studies, preprocessing approaches or even methods used to extract networks \citep[e.g.][]{Damoiseaux2006,Heuvel2008,Bellec2010c,Yeo2011,Power2011}. We however found that significant differences
in average connectivity existed between sites, as previously
reported by \cite{Yan2013a}. These site effects in connectivity may undermine the
generalization of the results derived at a single site. The inter-subject
(intra-site) standard deviation of the connections was found to be more than
twice as large as the inter-site absolute effect, on average across brain connections. This effect size measured in Cohen's d would be deemed small-to-moderate, which suggests that the impact of additive inter-site effects on statistical tests will be limited. This is a reassuring finding supporting the feasibility of statistical tests pooling fMRI data across multiple sites. Previous studies \citep{Sutton2008,Brown2011} had reported inter-site variance up to 10 times smaller than inter-subject variability, but these studies had much more homogeneous scanning environments than ours and also used different fMRI outcome measures. In our case, we still investigated only 3T scanners, mostly Siemens, and inter-site effects may be larger when considering other manufacturers or field strengths.\\

\paragraph{Statistical power and multisite rs-fMRI} After accounting for site-related additive effects in a GLM, the multisite simulation pooling 7 sites together showed detection power close to that of a monosite simulation with equivalent sample size. The difference was noticeable for small sample size (total $N=40$), and became very small for a sample size $N=120$. Another observation was that, for a given detection power, the lowest effect size that we were able to detect was more variable across connections for a low sample size. We demonstrated that a parametric group GLM does control precisely for the rate of false positive discoveries, even in multisite settings, as long as site covariates are included in the model. Taken together, these observations suggest to use sample sizes larger than $100$ subjects for GLM multisite studies. This conclusion may depend on the number of sites pooled in the study and the actual number of subjects in each of those sites, which we could not test in this work due to the size of the available sample.

\paragraph{Modeling site effects as random variables} We modeled the effect of each site on the average connectivity between any given pair of regions as a fixed effect. This means that the proposed GLM inference does apply only to collection of sites included in a given analysis. The linear mixed-effects model \citep{Chen2013} would allow more powerful inferences: by modeling site effects as random variables, following a specific distribution (e.g. Gaussian), we would be able to generalize observations potentially to any collection of sites, provided our assumptions are accurate. The sample of sites available for this study (7 at most) is however too small in our view to correctly estimate the variability of effects across sites. This work would also require to formulate and investigate empirically as well as on simulations different models for the distribution of inter-site variations of site effects (e.g. Gaussian distribution).

\paragraph{Site heteroscedasticity} We observed mild heteroscedasticity across sites. Our simulations showed that this does not compromise the control of false positive rate in the GLM, even under homoscedastic assumptions, with the range of contrasts we investigated. Regression models more robust to heteroscedasticity may be investigated in the future, e.g. weighted least squares regression or linear mixed-effects modeling \citep{Chen2013}.

\paragraph{Statistical power and sample size} For a medium effect size, e.g. $d=0.5$, the sensitivity was low (below $20\%$), even for monosite simulations with $N=120$ subjects. This sobering result supports the current trend in the literature to pool multiple data samples to increase sample size, at the cost of decreased homogeneity. We also found that resting-state studies based on 40 subjects or less, even at a single site, are seriously underpowered, except for extremely large effect sizes (Cohen's $d$ greater than 1.5). Finally, unbalanced patient allocation ratio in site samples greatly reduces sensitivity, even in monosite studies. Balanced datasets, i.e. with equal numbers of patients and controls at each site, should therefore be favored. 

\paragraph{Prediction}
Comparing the monosite and the multisite accuracy curves reveals a substantial
drop in accuracy from monosite to multisite across a broad range of effect
sizes. However, it should be noted that classifiers trained across multiple data
sources will likely generalize better to new observations, which is likely a
critical feature in most applications and reflects the true potential clinical
utility of this type of technique. Our conclusions are consistent with the work of
\cite{Nielsen2013}, which compares the prediction of a clinical diagnosis of
autism in monosite vs. multisite settings. The authors concluded that the
prediction accuracy for the multisite sample was significantly smaller than for
the monosite sample. A somewhat surprising observation in our analysis was that linear
correction for site-specific effects did not improve accuracy of prediction using
SVM. The SVM model seems to learn features that are invariant across sites,
maybe focusing on connections with the smallest site effect, or looking at
differences between connections similarly impacted by a site effect. Finally, an
important conclusion of our simulations was that the volume of brain connections
affected by a disease impacts accuracy as much as the effect size per
connection. This suggests that feature reduction and/or selection is a very
important step to improve sensitivity to small effect sizes.

\paragraph{Beyond additive site effect} An important limitation to our study is that we only investigated the impact of additive effects in brain connectivity across sites. Areas of future work include interactions between site effects and pathology, possibly in the form of polynomial and non-linear interactions. We hope that, in the future, fMRI data acquired on clinical cohorts at tens of sites will become available, which will enable researchers to test empirically the presence of such interaction effects. 

\paragraph{Other types of multisite data} Another limitation of our study is that we only investigated multisite data featuring roughly equal sample sizes with fairly balanced patient allocation ratios at each site. Multisite studies including a very large number of sites with sometimes only a few subjects per site are however quite common, e.g. the Alzheimer's disease neuroimaging initiative (ADNI) \citep{Mueller2005} and many pharmaceutical clinical trials at phase II and III \footnote{\url{http://www.rochetrials.com/trialDetailsGet.action?studyNumber=BP28248}}. In this type of design, the multisite effect may play a much more pronounced role than in our simulations as it cannot be modeled in the GLM, and will become an intrinsic added source of inter-subject variance \citep{Feaster2011}. Unfortunately, this type of design could not be tested with the current dataset due to the limited number of sites available. This represents an important avenue of future work. 

\paragraph{Underlying causes of the site effects}
Not all sites seemed to be equally impacted by the site effects, with sites like Berlin or Saint-Louis showing a small number of connections significantly different then the grand average connectivity matrix, while sites like Baltimore, Queensland and Oxford showed many more connections affected by the site effects. Interestingly this can potentially be due to temporal variance of the connections (see Supplementary Figure \ref{fig_std_ts_distribution}) partly explained by the scanner make since Queensland and Baltimore site used scanners from different makers (namely Bruker and Philips) than the rest of the sites used in this study (Siemens scanners). This may suggest that scanners SNR (signal to noise ratio) may partly explain the variance of connectivity. These differences may not be statistically significant, or they may reflect real differences due to protocol, scanner characteristics at these sites or differences in sampling across sites. Multiple causes may be interacting together to produce the site effects, as reported by \cite{Yan2013a}, although some of these sources of variance could be better controlled like the scanner parameters, paired with the use of a phantom to promote more homogeneous configurations across sites \citep{Friedman2006,Friedman2006a,Glover2012,Friedman2008}. Even in standardized experiments, it should be noted that differences in scanner protocols remain \citep{Brown2011}. A much larger multisite sample with systematically varying parameters could enable a data-driven identification of the critical parameters impacting site effects. The various releases made by the INDI initiative may fill that gap in the literature in the future, as the scanner protocols are much better described in recent releases, such as CoRR \citep{Zuo2014-ec}, than they were in the initial FCP release. These findings stress the need for more work to find the source of that variance rather than ad-hoc procedures to correct for them.

\section{Acknowledgments}
Parts of this work were presented at the 2013 annual meeting of the Organization for Human Brain Mapping, as well as the 2013 Alzheimer's Association International Conference (AAIC) \citep{Dansereau2013b}. The authors are grateful to the members of the 1000 functional connectome consortium for publicly releasing their datasets. The computational resources used to perform the data analysis were provided by ComputeCanada\footnote{\url{https://computecanada.org/}} and CLUMEQ\footnote{\url{http://www.clumeq.mcgill.ca/}}, which is funded in part by NSERC (MRS), FQRNT, and McGill University. This project was funded by NSERC grant number RN000028 and the Canadian
Consortium on Neurodegeneration in Aging (CCNA), through a grant from
the Canadian Institute of Health Research and funding from several partners including SANOFI-ADVENTIS R\&D. PB is supported by a salary award from ``Fonds de recherche du Qu\'ebec -- Sant\'e'' and the Courtois Foundation.

\section*{References}

\bibliographystyle{elsarticle-harv}
\bibliography{cdansereau}

\begin{thebibliography}{57}
\expandafter\ifx\csname natexlab\endcsname\relax\def\natexlab#1{#1}\fi
\expandafter\ifx\csname url\endcsname\relax
  \def\url#1{\texttt{#1}}\fi
\expandafter\ifx\csname urlprefix\endcsname\relax\def\urlprefix{URL }\fi

\bibitem[{Abraham et~al.(2016)Abraham, Milham, Di~Martino, Craddock, Samaras,
  Thirion, and Varoquaux}]{Abraham2016}
Abraham, A., Milham, M., Di~Martino, A., Craddock, R.~C., Samaras, D., Thirion,
  B., Varoquaux, G., 2016. Deriving reproducible biomarkers from multi-site
  resting-state data: An autism-based example. NeuroImage.

\bibitem[{Ad-Dab'bagh et~al.(2006)Ad-Dab'bagh, Einarson, Lyttelton, Muehlboeck,
  Mok, Ivanov, Vincent, Lepage, Lerch, Fombonne, and Evans}]{Ad-Dab'bagh2006}
Ad-Dab'bagh, Y., Einarson, D., Lyttelton, O., Muehlboeck, J.~S., Mok, K.,
  Ivanov, O., Vincent, R.~D., Lepage, C., Lerch, J., Fombonne, E., Evans,
  A.~C., 2006. The {CIVET} {Image-Processing} environment: A fully automated
  comprehensive pipeline for anatomical neuroimaging research. In: Corbetta, M.
  (Ed.), Proceedings of the 12th Annual Meeting of the Human Brain Mapping
  Organization. Neuroimage, Florence, Italy.

\bibitem[{Anand et~al.(2007)Anand, Li, Wang, Gardner, and Lowe}]{Anand2007}
Anand, A., Li, Y., Wang, Y., Gardner, K., Lowe, M.~J., 2007. Reciprocal effects
  of antidepressant treatment on activity and connectivity of the mood
  regulating circuit: an fmri study. The Journal of neuropsychiatry and
  clinical neurosciences.

\bibitem[{Bellec et~al.(2010{\natexlab{a}})Bellec, Carbonell, Perlbarg, and
  Evans}]{Bellec2010}
Bellec, P., Carbonell, F., Perlbarg, V., Evans, A.~C., 2010{\natexlab{a}}. {A
  neuroimaging analysis kit for Octave and Matlab}.
\newline\urlprefix\url{http://code.google.com/p/niak/}

\bibitem[{Bellec et~al.(2011)Bellec, Carbonell, Perlbarg, Lepage, Lyttelton,
  Fonov, Janke, Tohka, and Evans}]{Bellec2011}
Bellec, P., Carbonell, F.~M., Perlbarg, V., Lepage, C., Lyttelton, O., Fonov,
  V., Janke, A., Tohka, J., Evans, A.~C., 2011. {A neuroimaging analysis kit
  for Matlab and Octave}. In: Proceedings of the 17th International Conference
  on Functional Mapping of the Human Brain. pp. In Press+.

\bibitem[{Bellec et~al.(2012)Bellec, Lavoie-Courchesne, Dickinson, Lerch,
  Zijdenbos, and Evans}]{Bellec2012}
Bellec, P., Lavoie-Courchesne, S., Dickinson, P., Lerch, J.~P., Zijdenbos,
  A.~P., Evans, A.~C., 2012. {The pipeline system for Octave and Matlab (PSOM):
  a lightweight scripting framework and execution engine for scientific
  workflows.} Frontiers in neuroinformatics 6.
\newline\urlprefix\url{http://dx.doi.org/10.3389/fninf.2012.00007}

\bibitem[{Bellec et~al.(2010{\natexlab{b}})Bellec, Rosa-Neto, Lyttelton,
  Benali, and Evans}]{Bellec2010c}
Bellec, P., Rosa-Neto, P., Lyttelton, O.~C., Benali, H., Evans, A.~C., Jul.
  2010{\natexlab{b}}. {Multi-level bootstrap analysis of stable clusters in
  resting-state fMRI.} NeuroImage 51~(3), 1126--1139.
\newline\urlprefix\url{http://dx.doi.org/10.1016/j.neuroimage.2010.02.082}

\bibitem[{Benjamini and Hochberg(1995)}]{Benjamini1995}
Benjamini, Y., Hochberg, Y., 1995. {Controlling the false-discovery rate: a
  practical and powerful approach to multiple testing.} J. Roy. Statist. Soc.
  Ser. B 57, 289--300.

\bibitem[{Biswal et~al.(2010)Biswal, Mennes, Zuo, Gohel, Kelly, Smith,
  Beckmann, Adelstein, Buckner, Colcombe, Dogonowski, Ernst, Fair, Hampson,
  Hoptman, Hyde, Kiviniemi, K\"{o}tter, Li, Lin, Lowe, Mackay, Madden, Madsen,
  Margulies, Mayberg, McMahon, Monk, Mostofsky, Nagel, Pekar, Peltier,
  Petersen, Riedl, Rombouts, Rypma, Schlaggar, Schmidt, Seidler, Siegle, Sorg,
  Teng, Veijola, Villringer, Walter, Wang, Weng, Whitfield-Gabrieli,
  Williamson, Windischberger, Zang, Zhang, Castellanos, and
  Milham}]{Biswal2010}
Biswal, B.~B., Mennes, M., Zuo, X.-N.~N., Gohel, S., Kelly, C., Smith, S.~M.,
  Beckmann, C.~F., Adelstein, J.~S., Buckner, R.~L., Colcombe, S., Dogonowski,
  A.-M.~M., Ernst, M., Fair, D., Hampson, M., Hoptman, M.~J., Hyde, J.~S.,
  Kiviniemi, V.~J., K\"{o}tter, R., Li, S.-J.~J., Lin, C.-P.~P., Lowe, M.~J.,
  Mackay, C., Madden, D.~J., Madsen, K.~H., Margulies, D.~S., Mayberg, H.~S.,
  McMahon, K., Monk, C.~S., Mostofsky, S.~H., Nagel, B.~J., Pekar, J.~J.,
  Peltier, S.~J., Petersen, S.~E., Riedl, V., Rombouts, S.~A., Rypma, B.,
  Schlaggar, B.~L., Schmidt, S., Seidler, R.~D., Siegle, G.~J., Sorg, C., Teng,
  G.-J.~J., Veijola, J., Villringer, A., Walter, M., Wang, L., Weng, X.-C.~C.,
  Whitfield-Gabrieli, S., Williamson, P., Windischberger, C., Zang, Y.-F.~F.,
  Zhang, H.-Y.~Y., Castellanos, F.~X., Milham, M.~P., Mar. 2010. {Toward
  discovery science of human brain function.} Proceedings of the National
  Academy of Sciences of the United States of America 107~(10), 4734--4739.
\newline\urlprefix\url{http://dx.doi.org/10.1073/pnas.0911855107}

\bibitem[{Brown et~al.(2011)Brown, Mathalon, Stern, Ford, Mueller, Greve,
  McCarthy, Voyvodic, Glover, Diaz, et~al.}]{Brown2011}
Brown, G.~G., Mathalon, D.~H., Stern, H., Ford, J., Mueller, B., Greve, D.~N.,
  McCarthy, G., Voyvodic, J., Glover, G., Diaz, M., et~al., 2011. Multisite
  reliability of cognitive bold data. Neuroimage 54~(3), 2163--2175.

\bibitem[{Caramanos et~al.(2010)Caramanos, Fonov, Francis, Narayanan, Pike,
  Collins, and Arnold}]{Caramanos2010}
Caramanos, Z., Fonov, V.~S., Francis, S.~J., Narayanan, S., Pike, G.~B.,
  Collins, D.~L., Arnold, D.~L., 2010. Gradient distortions in mri:
  Characterizing and correcting for their effects on siena-generated measures
  of brain volume change. NeuroImage 49~(2), 1601--1611.

\bibitem[{Chang and Lin(2011)}]{Chang2011}
Chang, C.-C., Lin, C.-J., 2011. {LIBSVM}: A library for support vector
  machines. ACM Transactions on Intelligent Systems and Technology 2,
  27:1--27:27, software available at
  \url{http://www.csie.ntu.edu.tw/~cjlin/libsvm}.

\bibitem[{Chen et~al.(2013)Chen, Saad, Britton, Pine, and Cox}]{Chen2013}
Chen, G., Saad, Z.~S., Britton, J.~C., Pine, D.~S., Cox, R.~W., 2013. Linear
  mixed-effects modeling approach to fmri group analysis. Neuroimage 73,
  176--190.

\bibitem[{Cheng et~al.(2015)Cheng, Palaniyappan, Li, Kendrick, Zhang, Luo, Liu,
  Yu, Deng, Wang, Ma, Guo, Francis, Liddle, Mayer, Schumann, Li, and
  Feng}]{Cheng2015}
Cheng, W., Palaniyappan, L., Li, M., Kendrick, K.~M., Zhang, J., Luo, Q., Liu,
  Z., Yu, R., Deng, W., Wang, Q., Ma, X., Guo, W., Francis, S., Liddle, P.,
  Mayer, A.~R., Schumann, G., Li, T., Feng, J., May 2015. Voxel-based,
  brain-wide association study of aberrant functional connectivity in
  schizophrenia implicates thalamocortical circuitry. Npj Schizophrenia 1, --.
\newline\urlprefix\url{http://dx.doi.org/10.1038/npjschz.2015.16}

\bibitem[{Cohen(1992)}]{Cohen1992}
Cohen, J., 1992. A power primer. Psychological bulletin 112~(1), 155.

\bibitem[{Collins et~al.(1994)Collins, Neelin, Peters, and Evans}]{Collins1994}
Collins, D.~L., Neelin, P., Peters, T.~M., Evans, A.~C., 1994. {Automatic 3D
  intersubject registration of MR volumetric data in standardized Talairach
  space.} Journal of computer assisted tomography 18~(2), 192--205.
\newline\urlprefix\url{http://view.ncbi.nlm.nih.gov/pubmed/8126267}

\bibitem[{Cortes and Vapnik(1995)}]{Cortes1995}
Cortes, C., Vapnik, V., Sep. 1995. {Support-vector networks}. Machine Learning
  20~(3), 273--297.
\newline\urlprefix\url{http://dx.doi.org/10.1007/BF00994018}

\bibitem[{Damoiseaux et~al.(2006)Damoiseaux, Rombouts, Barkhof, Scheltens,
  Stam, Smith, and Beckmann}]{Damoiseaux2006}
Damoiseaux, J.~S., Rombouts, S. A. R.~B., Barkhof, F., Scheltens, P., Stam,
  C.~J., Smith, S.~M., Beckmann, C.~F., Sep. 2006. {Consistent resting-state
  networks across healthy subjects}. Proceedings of the National Academy of
  Sciences 103~(37), 13848--13853.
\newline\urlprefix\url{http://dx.doi.org/10.1073/pnas.0601417103}

\bibitem[{Dansereau et~al.(2014)Dansereau, Bellec, Lee, Pittau, Gotman, and
  Grova}]{Dansereau2014}
Dansereau, C., Bellec, P., Lee, K., Pittau, F., Gotman, J., Grova, C., 2014.
  Detection of abnormal resting-state networks in individual patients suffering
  from focal epilepsy: An initial step toward individual connectivity
  assessment. Frontiers in Neuroscience 8~(419).
\newline\urlprefix\url{http://www.frontiersin.org/brain_imaging_methods/10.3389/fnins.2014.00419/abstract}

\bibitem[{Dansereau et~al.(2013)Dansereau, Risterucci, Pich, Arnold, and
  Bellec}]{Dansereau2013b}
Dansereau, C., Risterucci, C., Pich, E.~M., Arnold, D., Bellec, P., 2013. A
  power analysis for multisite studies in resting-state functional
  connectivity, with an application to clinical trials in alzheimer's disease.
  Vol.~9. pp. P248 -- P249, alzheimer's Association International Conference
  2013 Alzheimer's Association International Conference 2013.
\newline\urlprefix\url{http://www.sciencedirect.com/science/article/pii/S1552526013011461}

\bibitem[{Desmond and Glover(2002)}]{Desmond2002}
Desmond, J., Glover, G., Aug. 2002. Estimating sample size in functional mri
  (fmri) neuroimaging studies: Statistical power analyses. Journal of
  Neuroscience Methods 118~(2), 115--128.
\newline\urlprefix\url{http://dx.doi.org/10.1016/s0165-0270(02)00121-8}

\bibitem[{Durnez et~al.(2014)Durnez, Moerkerke, and Nichols}]{Durnez2014}
Durnez, J., Moerkerke, B., Nichols, T.~E., 2014. Post-hoc power estimation for
  topological inference in fmri. Neuroimage 84, 45--64.

\bibitem[{Edward et~al.(2000)Edward, Windischberger, Cunnington, Erdler,
  Lanzenberger, Mayer, Endl, and Beisteiner}]{Edward2000}
Edward, V., Windischberger, C., Cunnington, R., Erdler, M., Lanzenberger, R.,
  Mayer, D., Endl, W., Beisteiner, R., Nov 2000. Quantification of fmri
  artifact reduction by a novel plaster cast head holder. Hum Brain Mapp
  11~(3), 207--213.

\bibitem[{Elliott et~al.(1999)Elliott, Bowtell, and Morris}]{Elliott1999}
Elliott, M.~R., Bowtell, R.~W., Morris, P.~G., Jun 1999. The effect of scanner
  sound in visual, motor, and auditory functional mri. Magn Reson Med 41~(6),
  1230--1235.

\bibitem[{Fair et~al.(2012)Fair, Nigg, Iyer, Bathula, Mills, Dosenbach,
  Schlaggar, Mennes, Gutman, Bangaru, Buitelaar, Dickstein, Martino, Kennedy,
  Kelly, Luna, Schweitzer, Velanova, Wang, Mostofsky, Castellanos, and
  Milham}]{Fair2012}
Fair, D.~A., Nigg, J.~T., Iyer, S., Bathula, D., Mills, K.~L., Dosenbach, N.
  U.~F., Schlaggar, B.~L., Mennes, M., Gutman, D., Bangaru, S., Buitelaar,
  J.~K., Dickstein, D.~P., Martino, A.~D., Kennedy, D.~N., Kelly, C., Luna, B.,
  Schweitzer, J.~B., Velanova, K., Wang, Y.-F., Mostofsky, S., Castellanos,
  F.~X., Milham, M.~P., 2012. Distinct neural signatures detected for adhd
  subtypes after controlling for micro-movements in resting state functional
  connectivity mri data. Front Syst Neurosci 6, 80.
\newline\urlprefix\url{http://dx.doi.org/10.3389/fnsys.2012.00080}

\bibitem[{Feaster et~al.(2011)Feaster, Mikulich-Gilbertson, and
  Brincks}]{Feaster2011}
Feaster, D., Mikulich-Gilbertson, S., Brincks, A., Sep. 2011. Modeling site
  effects in the design and analysis of multi-site trials. The American journal
  of drug and alcohol abuse 37~(5), 383--391.
\newline\urlprefix\url{http://dx.doi.org/10.3109/00952990.2011.600386}

\bibitem[{Fonov et~al.(2011)Fonov, Evans, Botteron, Almli, McKinstry, Collins,
  and {Brain Development Cooperative Group}}]{Fonov2011}
Fonov, V., Evans, A.~C., Botteron, K., Almli, C.~R., McKinstry, R.~C., Collins,
  D.~L., {Brain Development Cooperative Group}, Jan. 2011. {Unbiased average
  age-appropriate atlases for pediatric studies.} NeuroImage 54~(1), 313--327.
\newline\urlprefix\url{http://dx.doi.org/10.1016/j.neuroimage.2010.07.033}

\bibitem[{Friedman and Glover(2006)}]{Friedman2006a}
Friedman, L., Glover, G., Jun. 2006. Report on a multicenter fmri quality
  assurance protocol. Journal of magnetic resonance imaging : JMRI 23~(6),
  827--839.
\newline\urlprefix\url{http://dx.doi.org/10.1002/jmri.20583}

\bibitem[{Friedman et~al.(2006)Friedman, Glover, and Consortium}]{Friedman2006}
Friedman, L., Glover, G., Consortium, T.~F., Nov. 2006. Reducing interscanner
  variability of activation in a multicenter fmri study: Controlling for
  signal-to-fluctuation-noise-ratio (sfnr) differences. NeuroImage 33~(2),
  471--481.
\newline\urlprefix\url{http://dx.doi.org/10.1016/j.neuroimage.2006.07.012}

\bibitem[{Friedman et~al.(2008)Friedman, Stern, Brown, Mathalon, Turner,
  Glover, Gollub, Lauriello, Lim, Cannon, et~al.}]{Friedman2008}
Friedman, L., Stern, H., Brown, G.~G., Mathalon, D.~H., Turner, J., Glover,
  G.~H., Gollub, R.~L., Lauriello, J., Lim, K.~O., Cannon, T., et~al., 2008.
  Test--retest and between-site reliability in a multicenter fmri study. Human
  brain mapping 29~(8), 958--972.

\bibitem[{Giove et~al.(2009)Giove, Gili, Iacovella, Macaluso, and
  Maraviglia}]{Giove2009}
Giove, F., Gili, T., Iacovella, V., Macaluso, E., Maraviglia, B., Oct. 2009.
  {Images-based suppression of unwanted global signals in resting-state
  functional connectivity studies.} Magnetic resonance imaging 27~(8),
  1058--1064.
\newline\urlprefix\url{http://dx.doi.org/10.1016/j.mri.2009.06.004}

\bibitem[{Glover et~al.(2012)Glover, Mueller, Turner, van Erp, Liu, Greve,
  Voyvodic, Rasmussen, Brown, Keator, et~al.}]{Glover2012}
Glover, G.~H., Mueller, B.~A., Turner, J.~A., van Erp, T.~G., Liu, T.~T.,
  Greve, D.~N., Voyvodic, J.~T., Rasmussen, J., Brown, G.~G., Keator, D.~B.,
  et~al., 2012. Function biomedical informatics research network
  recommendations for prospective multicenter functional mri studies. Journal
  of Magnetic Resonance Imaging 36~(1), 39--54.

\bibitem[{Greicius et~al.(2004)Greicius, Srivastava, Reiss, and
  Menon}]{Greicius2004}
Greicius, M.~D., Srivastava, G., Reiss, A.~L., Menon, V., Mar. 2004.
  {Default-mode network activity distinguishes Alzheimer's disease from healthy
  aging: Evidence from functional MRI}. Proceedings of the National Academy of
  Sciences of the United States of America 101~(13), 4637--4642.
\newline\urlprefix\url{http://dx.doi.org/10.1073/pnas.0308627101}

\bibitem[{Hunter(2007)}]{matplotlib}
Hunter, J.~D., 2007. Matplotlib: A 2d graphics environment. Computing In
  Science \& Engineering 9~(3), 90--95.

\bibitem[{Jovicich et~al.(2016)Jovicich, Minati, Marizzoni, Marchitelli,
  Sala-Llonch, Bartr{\'e}s-Faz, Arnold, Benninghoff, Fiedler, Roccatagliata,
  et~al.}]{Jovicich2016}
Jovicich, J., Minati, L., Marizzoni, M., Marchitelli, R., Sala-Llonch, R.,
  Bartr{\'e}s-Faz, D., Arnold, J., Benninghoff, J., Fiedler, U., Roccatagliata,
  L., et~al., 2016. Longitudinal reproducibility of default-mode network
  connectivity in healthy elderly participants: A multicentric resting-state
  fmri study. NeuroImage 124, 442--454.

\bibitem[{Kilpatrick et~al.(2006)Kilpatrick, Zald, Pardo, and
  Cahill}]{Kilpatrick2006}
Kilpatrick, L., Zald, D., Pardo, J., Cahill, L., 2006. Sex-related differences
  in amygdala functional connectivity during resting conditions. Neuroimage
  30~(2), 452--461.

\bibitem[{Lund et~al.(2006)Lund, Madsen, Sidaros, Luo, and Nichols}]{Lund2006}
Lund, T.~E., Madsen, K.~H., Sidaros, K., Luo, W.-L., Nichols, T.~E., Jan. 2006.
  {Non-white noise in fMRI: does modelling have an impact?} NeuroImage 29~(1),
  54--66.
\newline\urlprefix\url{http://dx.doi.org/10.1016/j.neuroimage.2005.07.005}

\bibitem[{Milham et~al.(2012)Milham, Fair, Mennes, and Mostofsky}]{ADHD200}
Milham, M.~P., Fair, D., Mennes, M., Mostofsky, S.~H., 2012. The adhd-200
  consortium: a model to advance the translational potential of neuroimaging in
  clinical neuroscience. Frontiers in Systems Neuroscience 6~(62).
\newline\urlprefix\url{http://www.frontiersin.org/systems_neuroscience/10.3389/fnsys.2012.00062/full}

\bibitem[{Mueller et~al.(2005)Mueller, Weiner, Thal, Petersen, Jack, Jagust,
  Trojanowski, Toga, and Beckett}]{Mueller2005}
Mueller, S.~G., Weiner, M.~W., Thal, L.~J., Petersen, R.~C., Jack, C., Jagust,
  W., Trojanowski, J.~Q., Toga, A.~W., Beckett, L., Nov 2005. The alzheimer's
  disease neuroimaging initiative. Neuroimaging Clin N Am 15~(4), 869--77,
  xi--xii.
\newline\urlprefix\url{http://dx.doi.org/10.1016/j.nic.2005.09.008}

\bibitem[{Nielsen et~al.(2013)Nielsen, Zielinski, Fletcher, Alexander, Lange,
  Bigler, Lainhart, and Anderson}]{Nielsen2013}
Nielsen, J., Zielinski, B., Fletcher, T., Alexander, A., Lange, N., Bigler, E.,
  Lainhart, J., Anderson, J., 2013. Multisite functional connectivity mri
  classification of autism: Abide results. Frontiers in human neuroscience 7,
  --.
\newline\urlprefix\url{http://view.ncbi.nlm.nih.gov/pubmed/24093016}

\bibitem[{Orban et~al.(2015)Orban, Madjar, Savard, Dansereau, Tam, Das, Evans,
  Rosa-Neto, Breitner, and Bellec}]{Orban2015}
Orban, P., Madjar, C., Savard, M., Dansereau, C., Tam, A., Das, S., Evans,
  A.~C., Rosa-Neto, P., Breitner, J.~C., Bellec, P., oct 2015. Test-retest
  resting-state {fMRI} in healthy elderly persons with a family history of
  alzheimer's disease. Scientific Data 2, 150043.
\newline\urlprefix\url{http://dx.doi.org/10.1038/sdata.2015.43}

\bibitem[{Power et~al.(2012)Power, Barnes, Snyder, Schlaggar, and
  Petersen}]{Power2012}
Power, J.~D., Barnes, K.~A., Snyder, A.~Z., Schlaggar, B.~L., Petersen, S.~E.,
  Feb. 2012. {Spurious but systematic correlations in functional connectivity
  MRI networks arise from subject motion.} NeuroImage 59~(3), 2142--2154.
\newline\urlprefix\url{http://dx.doi.org/10.1016/j.neuroimage.2011.10.018}

\bibitem[{Power et~al.(2011)Power, Cohen, Nelson, Wig, Barnes, Church, Vogel,
  Laumann, Miezin, Schlaggar, and Petersen}]{Power2011}
Power, J.~D., Cohen, A.~L., Nelson, S.~M., Wig, G.~S., Barnes, K.~A., Church,
  J.~A., Vogel, A.~C., Laumann, T.~O., Miezin, F.~M., Schlaggar, B.~L.,
  Petersen, S.~E., Nov. 2011. {Functional Network Organization of the Human
  Brain}. Neuron 72~(4), 665--678.
\newline\urlprefix\url{http://dx.doi.org/10.1016/j.neuron.2011.09.006}

\bibitem[{Shehzad et~al.(2009)Shehzad, Kelly, Reiss, Gee, Gotimer, Uddin, Lee,
  Margulies, Roy, Biswal, Petkova, Castellanos, and Milham}]{Shehzad2009}
Shehzad, Z., Kelly, C.~M., Reiss, P.~T., Gee, D.~G., Gotimer, K., Uddin, L.~Q.,
  Lee, S. H.~H., Margulies, D.~S., Roy, A. K.~K., Biswal, B.~B., Petkova, E.,
  Castellanos, F.~X., Milham, M.~P., Oct. 2009. {The resting brain:
  unconstrained yet reliable.} Cerebral cortex (New York, N.Y. : 1991) 19~(10),
  2209--2229.
\newline\urlprefix\url{http://dx.doi.org/10.1093/cercor/bhn256}

\bibitem[{Sheline et~al.(2010)Sheline, Price, Yan, and Mintun}]{Sheline2010}
Sheline, Y.~I., Price, J.~L., Yan, Z., Mintun, M.~A., 2010. Resting-state
  functional mri in depression unmasks increased connectivity between networks
  via the dorsal nexus. Proceedings of the National Academy of Sciences
  107~(24), 11020--11025.

\bibitem[{Sutton et~al.(2008)Sutton, Goh, Hebrank, Welsh, Chee, and
  Park}]{Sutton2008}
Sutton, B.~P., Goh, J., Hebrank, A., Welsh, R.~C., Chee, M.~W., Park, D.~C.,
  2008. Investigation and validation of intersite fmri studies using the same
  imaging hardware. Journal of Magnetic Resonance Imaging 28~(1), 21--28.

\bibitem[{Tam et~al.(2015)Tam, Dansereau, Badhwar, Orban, Belleville, Chertkow,
  Dagher, Hanganu, Monchi, Rosa-Neto, Shmuel, Wang, Breitner, and
  Bellec}]{Tam2015}
Tam, A., Dansereau, C., Badhwar, A., Orban, P., Belleville, S., Chertkow, H.,
  Dagher, A., Hanganu, A., Monchi, O., Rosa-Neto, P., Shmuel, A., Wang, S.,
  Breitner, J., Bellec, P., 2015. Common effects of amnestic mild cognitive
  impairment on resting-state connectivity across four independent studies.
  Frontiers in Aging Neuroscience 7~(242).
\newline\urlprefix\url{http://www.frontiersin.org/aging_neuroscience/10.3389/fnagi.2015.00242/abstract}

\bibitem[{van~den Heuvel et~al.(2008)van~den Heuvel, Mandl, and
  Hulshoff~Pol}]{Heuvel2008}
van~den Heuvel, M., Mandl, R., Hulshoff~Pol, H., Apr. 2008. {Normalized Cut
  Group Clustering of Resting-State fMRI Data}. PLoS ONE 3~(4), e2001+.
\newline\urlprefix\url{http://dx.doi.org/10.1371/journal.pone.0002001}

\bibitem[{Van~Dijk et~al.(2010)Van~Dijk, Hedden, Venkataraman, Evans, Lazar,
  and Buckner}]{VanDijk2010}
Van~Dijk, K.~R., Hedden, T., Venkataraman, A., Evans, K.~C., Lazar, S.~W.,
  Buckner, R.~L., Jan. 2010. {Intrinsic functional connectivity as a tool for
  human connectomics: theory, properties, and optimization.} Journal of
  neurophysiology 103~(1), 297--321.
\newline\urlprefix\url{http://dx.doi.org/10.1152/jn.00783.2009}

\bibitem[{Van~Dijk et~al.(2012)Van~Dijk, Sabuncu, and Buckner}]{VanDijk2012}
Van~Dijk, K.~R., Sabuncu, M.~R., Buckner, R.~L., 2012. The influence of head
  motion on intrinsic functional connectivity mri. Neuroimage 59~(1), 431--438.

\bibitem[{Vanhoutte et~al.(2006)Vanhoutte, Verhoye, and der
  Linden}]{Vanhoutte2006}
Vanhoutte, G., Verhoye, M., der Linden, A.~V., May 2006. Changing body
  temperature affects the t2* signal in the rat brain and reveals hypothalamic
  activity. Magn Reson Med 55~(5), 1006--1012.
\newline\urlprefix\url{http://dx.doi.org/10.1002/mrm.20861}

\bibitem[{Wang et~al.(2012)Wang, Liu, Qin, Wang, Zhang, Jiang, and
  Yu}]{Wang2012}
Wang, D., Liu, B., Qin, W., Wang, J., Zhang, Y., Jiang, T., Yu, C., Dec. 2012.
  {KIBRA gene variants are associated with synchronization within the
  default-mode and executive control networks}. NeuroImage.
\newline\urlprefix\url{http://dx.doi.org/10.1016/j.neuroimage.2012.12.022}

\bibitem[{Worsley and Friston(1995)}]{Worsley1995}
Worsley, K.~J., Friston, K.~J., Sep. 1995. {Analysis of fMRI Time-Series
  Revisited---Again}. NeuroImage 2~(3), 173--181.
\newline\urlprefix\url{http://dx.doi.org/10.1006/nimg.1995.1023}

\bibitem[{Yan et~al.(2009)Yan, Liu, He, Zou, Zhu, Zuo, Long, and
  Zang}]{Yan2009}
Yan, C., Liu, D., He, Y., Zou, Q., Zhu, C., Zuo, X., Long, X., Zang, Y., May
  2009. {Spontaneous Brain Activity in the Default Mode Network Is Sensitive to
  Different Resting-State Conditions with Limited Cognitive Load}. PLoS ONE
  4~(5), e5743+.
\newline\urlprefix\url{http://dx.doi.org/10.1371/journal.pone.0005743}

\bibitem[{Yan et~al.(2013)Yan, Craddock, Zuo, Zang, and Milham}]{Yan2013a}
Yan, C.-G.~G., Craddock, C.~C., Zuo, X.-N.~N., Zang, Y.-F.~F., Milham, M.~P.,
  Oct. 2013. {Standardizing the intrinsic brain: towards robust measurement of
  inter-individual variation in 1000 functional connectomes.} NeuroImage 80,
  246--262.
\newline\urlprefix\url{http://view.ncbi.nlm.nih.gov/pubmed/23631983}

\bibitem[{Yeo et~al.(2011)Yeo, Krienen, Sepulcre, Sabuncu, Lashkari,
  Hollinshead, Roffman, Smoller, Z\"{o}llei, Polimeni, Fischl, Liu, and
  Buckner}]{Yeo2011}
Yeo, B.~T., Krienen, F.~M., Sepulcre, J., Sabuncu, M.~R., Lashkari, D.,
  Hollinshead, M., Roffman, J.~L., Smoller, J.~W., Z\"{o}llei, L., Polimeni,
  J.~R., Fischl, B., Liu, H., Buckner, R.~L., Sep. 2011. {The organization of
  the human cerebral cortex estimated by intrinsic functional connectivity.}
  Journal of neurophysiology 106~(3), 1125--1165.
\newline\urlprefix\url{http://dx.doi.org/10.1152/jn.00338.2011}

\bibitem[{Zuo et~al.(2014)Zuo, Anderson, Bellec, Birn, Biswal, Blautzik,
  Breitner, Buckner, Calhoun, Castellanos, Chen, Chen, Chen, Chen, Colcombe,
  Courtney, Craddock, Di~Martino, Dong, Fu, Gong, Gorgolewski, Han, He, He, Ho,
  Holmes, Hou, Huckins, Jiang, Jiang, Kelley, Kelly, King, LaConte, Lainhart,
  Lei, Li, Li, Li, Lin, Liu, Liu, Liu, Liu, Lu, Lu, Luna, Luo, Lurie, Mao,
  Margulies, Mayer, Meindl, Meyerand, Nan, Nielsen, O'Connor, Paulsen,
  Prabhakaran, Qi, Qiu, Shao, Shehzad, Tang, Villringer, Wang, Wang, Wei, Wei,
  Weng, Wu, Xu, Yang, Yang, Zang, Zhang, Zhang, Zhang, Zhang, Zhao, Zhen, Zhou,
  Zhu, and Milham}]{Zuo2014-ec}
Zuo, X.-N., Anderson, J.~S., Bellec, P., Birn, R.~M., Biswal, B.~B., Blautzik,
  J., Breitner, J. C.~S., Buckner, R.~L., Calhoun, V.~D., Castellanos, F.~X.,
  Chen, A., Chen, B., Chen, J., Chen, X., Colcombe, S.~J., Courtney, W.,
  Craddock, R.~C., Di~Martino, A., Dong, H.-M., Fu, X., Gong, Q., Gorgolewski,
  K.~J., Han, Y., He, Y., He, Y., Ho, E., Holmes, A., Hou, X.-H., Huckins, J.,
  Jiang, T., Jiang, Y., Kelley, W., Kelly, C., King, M., LaConte, S.~M.,
  Lainhart, J.~E., Lei, X., Li, H.-J., Li, K., Li, K., Lin, Q., Liu, D., Liu,
  J., Liu, X., Liu, Y., Lu, G., Lu, J., Luna, B., Luo, J., Lurie, D., Mao, Y.,
  Margulies, D.~S., Mayer, A.~R., Meindl, T., Meyerand, M.~E., Nan, W.,
  Nielsen, J.~A., O'Connor, D., Paulsen, D., Prabhakaran, V., Qi, Z., Qiu, J.,
  Shao, C., Shehzad, Z., Tang, W., Villringer, A., Wang, H., Wang, K., Wei, D.,
  Wei, G.-X., Weng, X.-C., Wu, X., Xu, T., Yang, N., Yang, Z., Zang, Y.-F.,
  Zhang, L., Zhang, Q., Zhang, Z., Zhang, Z., Zhao, K., Zhen, Z., Zhou, Y.,
  Zhu, X.-T., Milham, M.~P., 9~Dec. 2014. An open science resource for
  establishing reliability and reproducibility in functional connectomics. Sci
  Data 1, 140049.

\end{thebibliography}

\pagebreak

\clearpage
\appendix

\clearpage
\pagebreak
\renewcommand{\thefigure}{S\arabic{figure}}
\renewcommand{\thetable}{S\arabic{table}}
\setcounter{figure}{0}
\begin{center}
\emph{Supplementary Material {--} Statistical power and prediction accuracy in multisite resting-state fMRI connectivity}\\

\vspace{\baselineskip}Submitted to Neuroimage.\\

\vspace{\baselineskip}C. Dansereau$^{1,2}$,  Y. Benhajali$^{1,3}$,C. Risterucci$^{4}$, E. Merlo Pich$^{4}$, P. Orban$^{1}$, D. Arnold$^{5}$, P. Bellec$^{1,2}$\\

\end{center}

$^1$Centre de Recherche de l'Institut Universitaire de G\'eriatrie de Montr\'eal, Montr\'eal, CA\\
$^2$Department of Computer Science and Operations Research, University of Montreal, Montreal, CA\\
$^3$D\'epartement d'anthropologie, Universit\' de Montr\'eal, Montr\'eal, CA\\
$^4$Clinical Imaging, pRED, F.Hoffman-La Roche, Basel, CH\\
$^5$NeuroRx inc., Montr\'eal, CA\\

For all questions regarding the paper, please address correspondence to Pierre Bellec, CRIUGM, 4545 Queen Mary, Montreal, QC, H3W 1W5, Canada. Email: pierre.bellec (at) criugm.qc.ca.\\

\begin{figure}[htbp]
\centering
\includegraphics[width=0.75\textwidth]{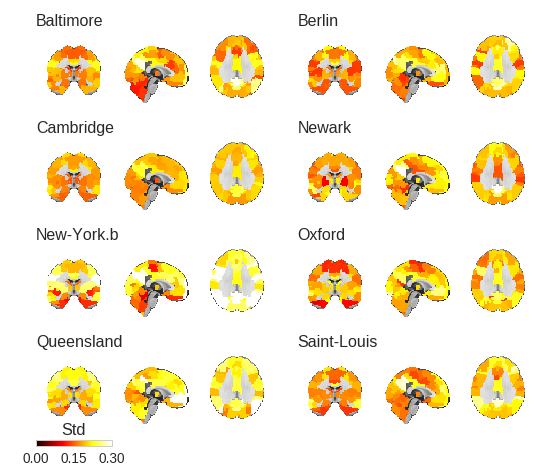}
\caption[]{
Standard deviation of resting-state connectivity across subjects, in the DMN, for each site, superimposed on the MNI152 template.
}
\label{fig_std_DMNs}
\end{figure}

\begin{figure}[htbp]
\centering
\includegraphics[width=0.50\textwidth]{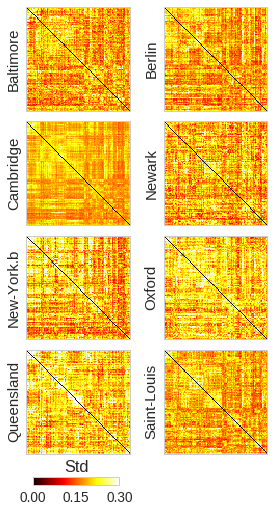}
\caption[]{
Standard deviation of resting-state connectivity across subjects, for the full connectome and each site.
}
\label{fig_std_connectomes}
\end{figure}

\begin{figure}[htbp]
\begin{center}
\includegraphics[width=0.75\linewidth]{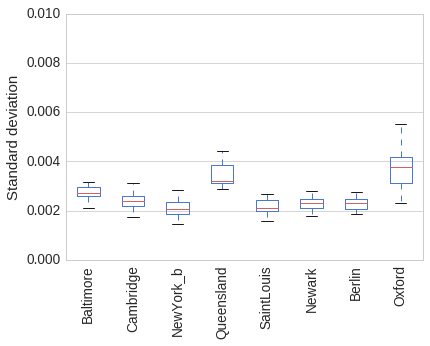}
\end{center}
\caption[]{
Standard deviation of resting-state time-series across subjects, averaged across all connections, at each site.
}
\label{fig_std_ts_distribution}
\end{figure}

\begin{figure}[htbp]
\begin{center}
\includegraphics[width=\linewidth]{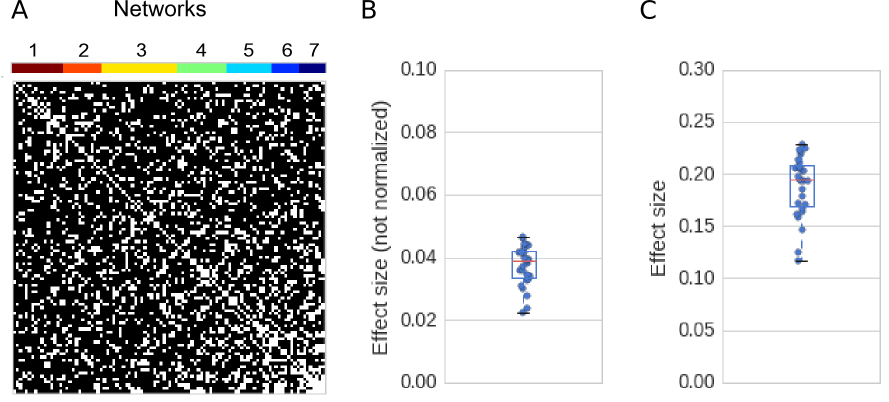}
\end{center}
\caption[]{
Panel A shows the results of a White test for homoscedasticity, across sites. Panel B show the average absolute difference in standard deviation between any pair of sites, and Panel C show the same difference, relative to the average of the standard deviation at the two sites. 
}
\label{fig_hetero}
\end{figure}

\begin{figure}[htbp]
\begin{center}
\includegraphics[width=\linewidth]{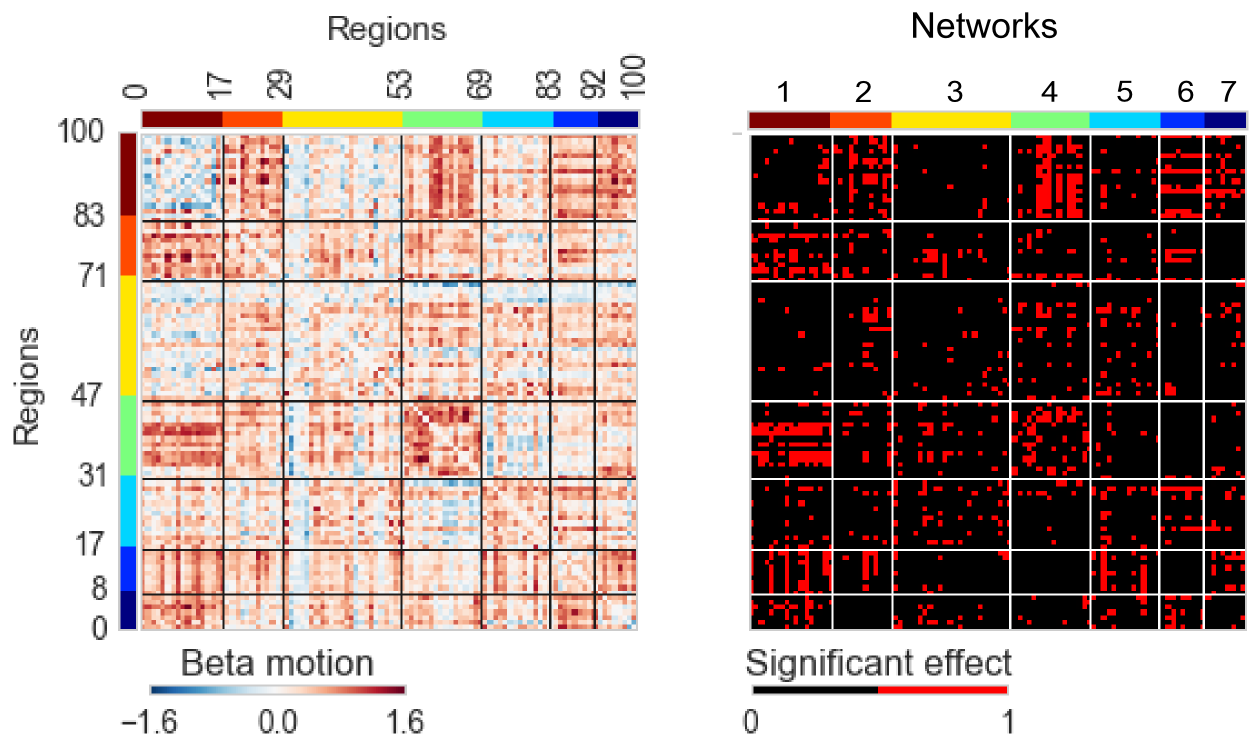}
\end{center}
\caption[Connectome variability across sites]{
The figure shows average connectomes across all sites, as well as connections with a significant motion effect.
}
\label{fig_connectome_variability_motion}
\end{figure}

\begin{figure}[htbp]
\begin{center}
\includegraphics[width=\linewidth]{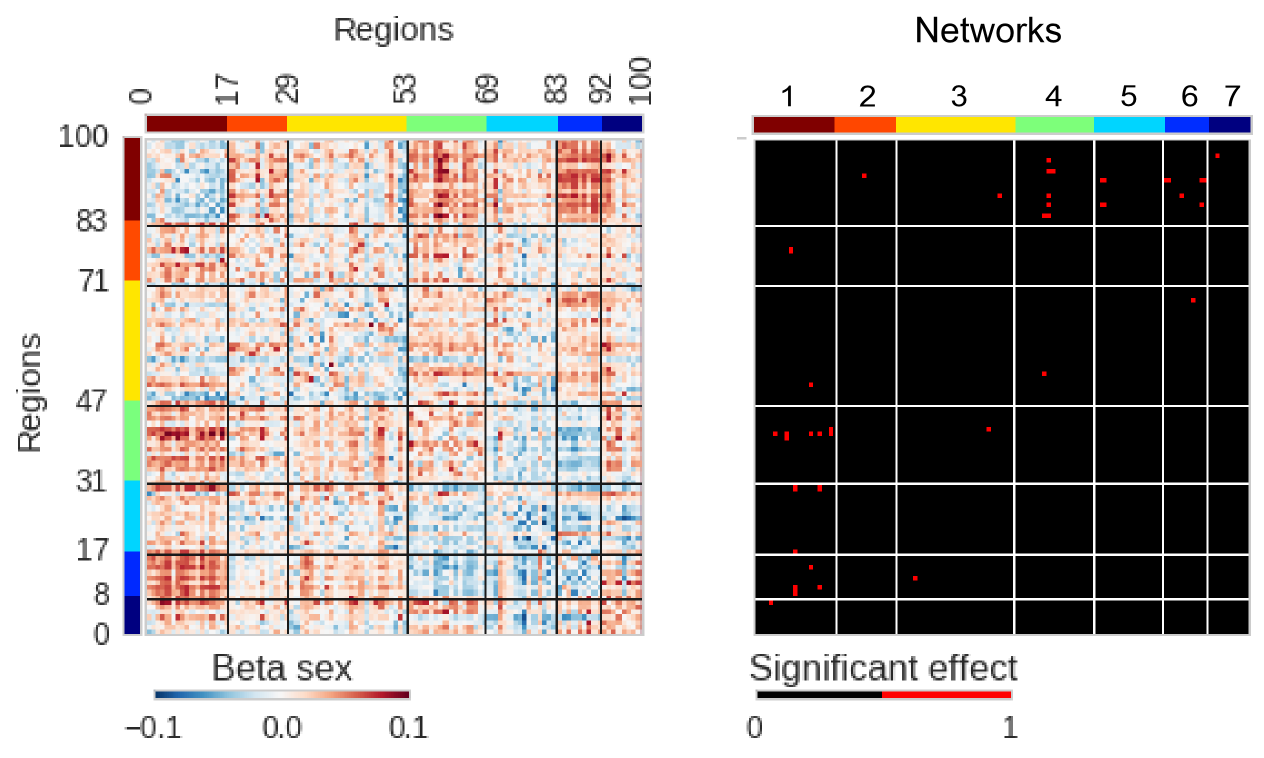}
\end{center}
\caption[Connectome variability across sites]{
The figure shows average connectomes across all sites, as well as connections with a significant sex effect.
}
\label{fig_connectome_variability_sex}
\end{figure}

\begin{figure}[htbp]
\begin{center}
\includegraphics[width=\linewidth]{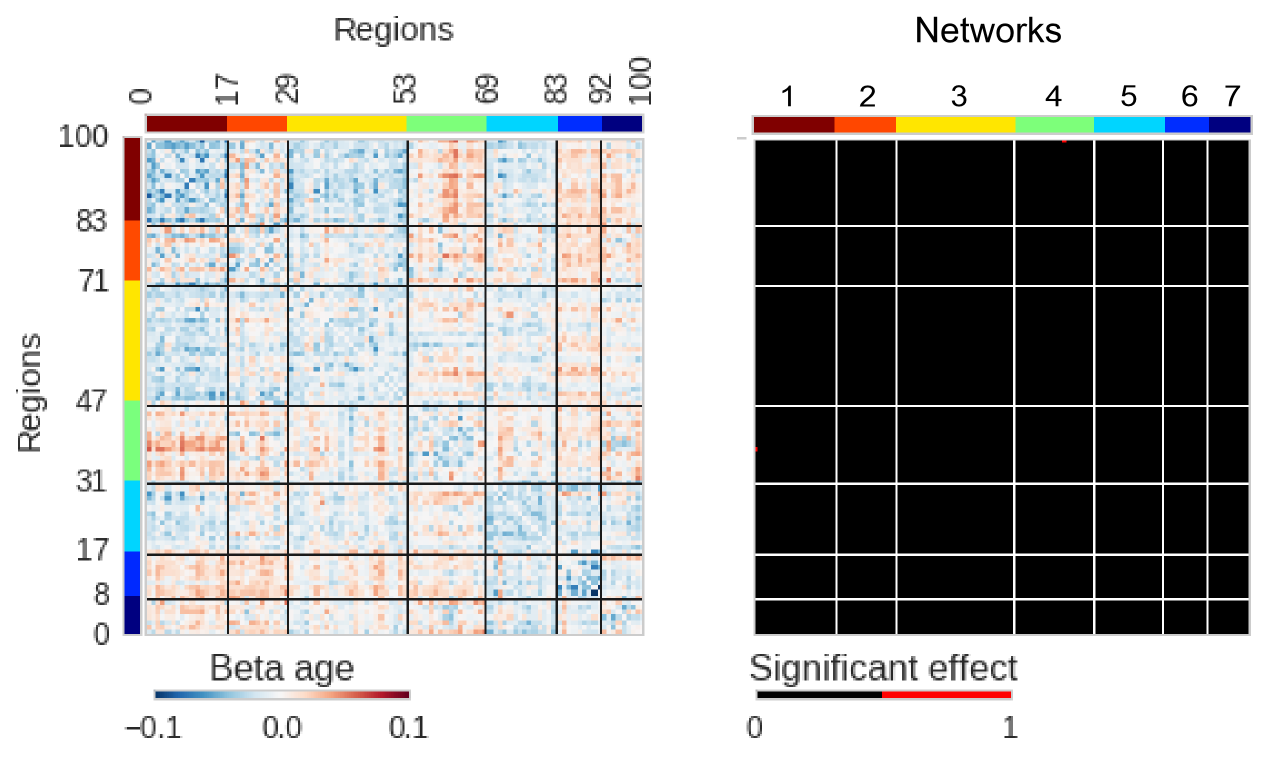}
\end{center}
\caption[Connectome variability across sites]{
The figure shows average connectomes across all sites, as well as connections with a significant age effect.
}
\label{fig_connectome_variability_age}
\end{figure}

\begin{figure}[htbp]
\begin{center}
\includegraphics[width=\linewidth]{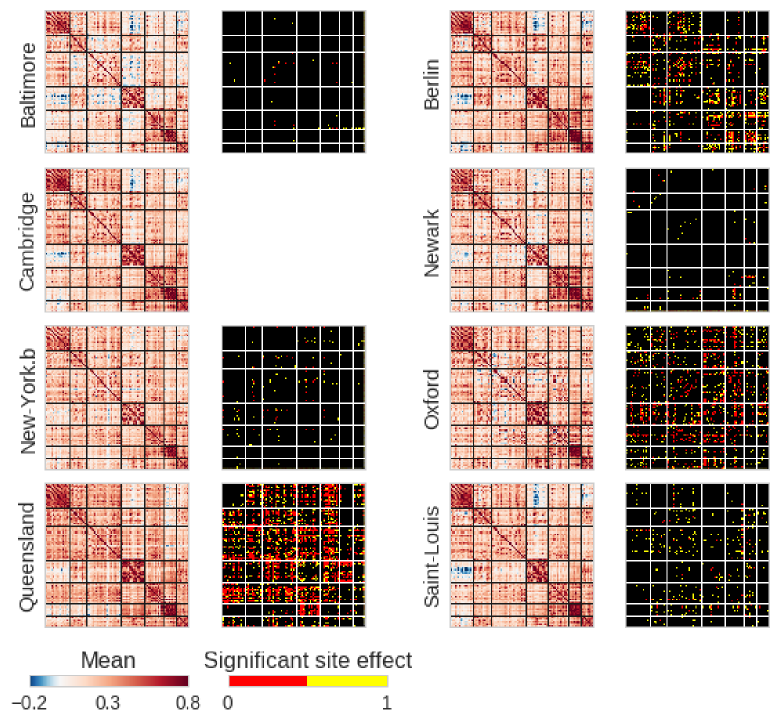}
\end{center}
\caption[]{
Average connectomes for individual sites, as well as connections with a significant site effect. This Figure is identical to Figure 2 in the paper, with the difference that the Cambridge site was excluded from the analysis. The intersection ($\cap$) of the significant site effects are shown in red and the symmetric difference ($\bigtriangleup$) of the significant site effects are shown in yellow. Baltimore $\cap:9,\bigtriangleup:16$, Berlin $\cap:318,\bigtriangleup:333$, Newark $\cap:23,\bigtriangleup:36$, New-York.b $\cap:25,\bigtriangleup:45$, Oxford $\cap:377,\bigtriangleup:251$, Queensland $\cap:946,\bigtriangleup:389$, Saint-Louis $\cap:49,\bigtriangleup:162$
}
\label{fig_connectome_variability_no_cambridge}
\end{figure}

\end{document}